\documentclass[a4paper,11pt]{article}

\usepackage{graphicx}
\usepackage{caption}
\usepackage{subcaption}
\usepackage{slashed,nicefrac}

\usepackage{mathtools}

\usepackage{fullpage}
\usepackage[T1]{fontenc}

\usepackage{amssymb,amsmath,cite}

\usepackage{xcolor,bbold,comment}

\usepackage{amsthm}

\def\S{\Sigma}

\def\p{\partial}
\def\ls{\left[}
\def\rs{\right]}

\newcommand{\be}{\begin{eqnarray}}
\newcommand{\ee}{\end{eqnarray}}
\allowdisplaybreaks

\usepackage[utf8]{inputenc}
\usepackage[english]{babel}

\numberwithin{equation}{section}
\begin{document}

\begin{titlepage}
	\thispagestyle{empty}
	\begin{flushright}
		
	\end{flushright}

	\vspace{35pt}
	
	\begin{center}
	    { \Large{\bf Complex Linear Multiplets 
	    and Local Supersymmetry Breaking}} 
		
		\vspace{50pt}
		
		{Fotis~Farakos$^{1,2}$, Alex~Kehagias$^{3}$ and Nikolaos~Liatsos$^{3}$}
		
		\vspace{25pt}

		$^1${\it  Dipartimento di Fisica ``Galileo Galilei''\\
		Universit\`a di Padova, Via Marzolo 8, 35131 Padova, Italy}
		
		\vspace{15pt}
		
	    $^2${\it   INFN, Sezione di Padova \\
		Via Marzolo 8, 35131 Padova, Italy}
		
		\vspace{15pt} 
		
        $^3${\it   Physics Division, National Technical University of Athens \\
        15780 Zografou Campus, Athens, Greece}
		
		\vspace{15pt}

		\vspace{40pt}
		
		{ABSTRACT} 
	\end{center}
	
We study supersymmetry breaking from a complex linear superfield coupled to 4D N=1 supergravity. 
The theory has two classically decoupled vacua, one supersymmetric and one with broken and intrinsically non-linear supersymmetry. 
Depending on the values of the parameters the scalar potential can lead to no-scale Minkwoski or a stable de Sitter or anti-de Sitter vacuum. 
We also provide a dual description of the system in terms of a nilpotent chiral superfield and a standard chiral coupled to supergravity.

\vspace{10pt}

\bigskip

\end{titlepage}


\newpage

\section{Introduction}

The two typical sources for positive contribution to the vacuum energy 
in 4D N=1 supergravity are the auxiliary fields of the chiral and the vector multiplets \cite{Wess:1992cp}. 
In particular, a positive contribution to the scalar potential is generated when in the presence of 
a superpotential the auxiliary fields of the chiral multiplets are integrated out. 
In addition, if there is a gauged isometry of the K\"ahler manifold, 
then a positive contribution to the scalar potential is generated from 
integrating out the auxiliary fields of the vector multiplets. 
These are also the same sources that are typically responsible for the breaking of supersymmetry. 
Chiral multiplets however are not the only possible candidates for the description of scalars 
and one can instead work with the so-called complex linear ones \cite{Gates:1980az,Deo:1985ix}. 
Such multiplets typically contain the same type of physical fields as the chiral multiplet, 
but have significantly different auxiliary fields. 
Since the breaking of supersymmetry is a topic that plays a key role in modern applications of supergravity 
it is important to understand its properties thoroughly for any kind of supermultiplet.

In this work we study the only known ghost-free mechanism where the scalar auxiliary field of a single complex linear multiplet 
is integrated out and gets a non-trivial vacuum-expectation-value (VEV). 
Therefore it is responsible for the breaking of supersymmetry and the generation of a positive contribution to the scalar potential. 
This construction has been studied in global supersymmetry \cite{Farakos:2013zsa,Farakos:2014iwa,Farakos:2015vba} 
and in linearized supergravity \cite{Koci:2016rqf}. 
Notably it has been observed in the early literature that supersymmetry breaking can occur 
from such a multiplet but it is treated only in global supersymmetry and it is also unclear if ghost states would exist or not in the model of \cite{Deo:1984cg}. 
In this work we follow the construction of \cite{Farakos:2013zsa} where ghost states are shown to be absent. 
However, there are other striking physical properties that do agree with the early findings of \cite{Deo:1984cg}: 
e.g. that a previously auxiliary spinor propagates only in the broken vacuum and becomes the Goldstone mode of supersymmetry.

Terms similar in spirit have also been constructed for other multiplets. 
For example in \cite{Cecotti:1986jy,Koehn:2012ar,Farakos:2012qu,BjarkeGudnason:2018aij} a similar setup for chiral multiplets is studied with focus 
on the possible supersymmetry breaking, 
whereas their non-trivial application for cosmology is discussed for example in \cite{Koehn:2013upa,Aoki:2014pna,Bielleman:2016grv,Yamada:2021kxv}, 
and their string theory origin is studied e.g. in \cite{Ciupke:2015msa,Deen:2017dpm}. 
In a similar vein (as we will show in section 3) there has also been a study on new types of Fayet--Iliopoulos terms that 
lead to a breaking of supersymmetry without requiring the gauging 
of the R-symmetry \cite{Cribiori:2017laj,Kuzenko:2018jlz,Aldabergenov:2018nzd,Antoniadis:2018oeh,Antoniadis:2019nwz}. 
The self-consistency of such terms, from the effective field theory perspective, has been scrutinized in \cite{Jang:2021fce,Jang:2021vpb}, 
and possible alternatives have been also studied in \cite{Farakos:2018sgq,Jang:2020cbe,Jang:2021xve} 
which relax some properties of supersymmetric scalar potentials. 
In addition, 
there exist modified (nilpotent or not) complex linear multiplets which can serve as the goldstino superfield \cite{Kuzenko:2011ti,Kuzenko:2015yxa}, 
however in those cases the auxiliary fields do not necessarily have a VEV. 
Similar constructions also exist for modified real linear multiplets \cite{Kuzenko:2017oni,Farakos:2018aml}.

Returning to the complex linear multiplet, 
the only scalar auxiliary field appears with a negative 
sign in the typical 2-derivative Lagrangian, 
that is \cite{Gates:1980az,Deo:1985ix} 
\be
{\cal L}_{kinetic} \, \sim \,  - F \overline F \, , 
\ee
instead of a positive one as in the chiral multiplet. 
Therefore, the presence of a ``superpotential'' for the complex linear would possibly lead to an inconsistent theory 
due to a ghost goldstino multiplet correlated to the potential negative energy contribution, 
and in any case it is not possible to include a superpotential for a pure complex linear due to the restrictions from supersymmetry. 
To this end the way to break supersymmetry with the complex linear is by introducing higher order terms such that the following term 
is generated \cite{Farakos:2013zsa} 
\be
{\cal L}_{higher \, order} \, \sim \,  + F^2 \overline F^2 \,. 
\ee
Here we go a step further in the understanding of this coupling by constructing the full 4D N=1 supergravity embedding and studying its properties.

The rest of the article is organized as follows: 
First we give a detailed analysis about the coupling of the complex linear superfield in supergravity in section 2 
(for standard and non-standard couplings as e.g. in \cite{Mahanthappa:1986cc}), 
and then we discuss the coupling of the superspace higher derivative term that leads to supersymmetry breaking in section 3. 
We find that the minimal model has two vacua that arise from the two different solutions to the equations of motion 
for the complex linear superfield auxiliary fields. 
Our component form study shows that one solution describes a supersymmetric background with a single scalar multiplet (we analyze this in section 4), 
whereas the other solution that exists only due to the higher order term describes a background with spontaneously broken supersymmetry that 
we analyze in section 3. 
Interestingly the background with broken supersymmetry is described with an intrinsically non-linear 
realization \cite{Lindstrom:1979kq,Kapustnikov:1981de,Samuel:1982uh,Farakos:2013ih,Bergshoeff:2015tra} 
and on top of that also contains a standard scalar multiplet (such models were studied e.g. in \cite{Hasegawa:2015bza}). 
We further verify our component form results by performing a superspace analysis in section 5.  
In section 6 we also perform a further analysis for non-standard complex linear couplings  
including also the higher order term we introduced in section 3. 
We give few concluding remarks in section 7.

\section{The complex linear multiplet in supergravity}

The complex linear multiplet and its coupling to 4D N=1 supergravity plays a central role in our work, 
therefore we will give a careful account of its properties in this section following throughout the conventions of \cite{Wess:1992cp}. 
In the so-called old-minimal supergravity, the complex linear (or non-minimal) 
multiplet is defined as \cite{Gates:1980az,Deo:1985ix}  
\be
\label{complin}
(\overline{\cal D}^2 -8 {\cal R}) \S=0 \, ,
\ee
where we have introduced the old-minimal supergravity chiral projection \cite{Wess:1992cp}. 
The component fields of the complex linear multiplet are defined as 
\be
\label{Scomp}
\begin{aligned}
A&=\Sigma| \,  ,\quad &&F=-\frac{1}{4} \mathcal{D}^2 \Sigma| \, , \\
\rho_\alpha&= \frac{1}{\sqrt{2}} \mathcal{D}_\alpha \overline \Sigma| \, , \quad &&\lambda_\alpha=\frac{1}{\sqrt{2}} \mathcal{D}_\alpha \Sigma | \, , \\
P_{\alpha \dot{\beta}}&=\overline{\mathcal{D}}_{\dot{\beta}} \mathcal{D}_\alpha \Sigma| \, , \quad &&{\overline P}_{\beta \dot{\alpha}} =-\mathcal{D}_\beta \overline{\mathcal{D}}_{\dot{\alpha}} \overline\Sigma| \, , \\
\chi_\alpha &= \frac{1}{2} \overline{\mathcal{D}}_{\dot{\alpha}} \mathcal{D}_\alpha \overline{\mathcal{D}}^{\dot{\alpha}} \overline \Sigma| \, , \quad &&\overline{\chi}_{\dot{\alpha}}  =\frac{1}{2}  \mathcal{D}^\alpha \overline{\mathcal{D}}_{\dot{\alpha}} \mathcal{D}_\alpha \Sigma| \, .
\end{aligned}
\ee
Under a local supersymmetry transformation with Grassmann parameters $\zeta_\alpha$ and $\overline{\zeta}_{\dot{\alpha}}= (\zeta_\alpha)^*$, 
the fermionic component fields of $\Sigma$ transform as\footnote{The full expression for the supergravity transformation of the auxiliary fermion $\chi_\alpha$ is given in the appendix.} 
\begin{align}
 \delta \lambda_\alpha= & - \sqrt{2} \zeta_\alpha F + \frac{1}{\sqrt{2}} P_{\alpha \dot{\beta}} \overline{\zeta}^{\dot{\beta}}   \, , \\
 \delta \rho_\alpha  = & - \frac{\sqrt{2}}{3} \zeta_\alpha \overline M \overline A - i \sqrt{2}  (\sigma^m \overline\zeta )_\alpha \hat{D}_m \overline A + \frac{1}{\sqrt{2}} {\overline P}_{\alpha \dot{\beta}} \overline\zeta^{\dot{\beta}} \, , \\
 \nonumber \delta \chi_\alpha = & -2i ( \sigma^b \overline\sigma^a \zeta )_\alpha \hat{D}_a {\overline P}_b - i ( \sigma^a \overline\sigma^b \zeta  )_\alpha \hat{D}_a {\overline P}_b - 4 ( \sigma^b \overline\sigma^a \zeta )_\alpha \hat{D}_a \hat{D}_b \overline A \\ & \nonumber - \frac{8}{3} \zeta_\alpha \overline M \overline F - ( \sigma^m \overline \sigma^n \zeta )_\alpha b_m {\overline P}_n - \frac{1}{3} ( \sigma^n \overline \sigma^m \zeta )_\alpha b_m {\overline P}_n + \frac{2}{3} \zeta_\alpha \overline A R \\
 & - \frac{8i}{3} \zeta_\alpha b^m \hat{D}_m \overline A - \frac{4i}{3} \zeta_\alpha \overline A {e_a}^m \mathcal{D}_m b^a - \frac{4}{9} \zeta_\alpha \overline A b_m b^m   \\
 \nonumber & - 2 i ( \sigma^m \overline \zeta )_\alpha \hat{D}_m \overline F + \frac{1}{3}  ( \sigma^m \overline \zeta )_\alpha M \overline P_m - \frac{4}{3}  ( \sigma^m \overline \zeta )_\alpha b_m \overline F \\
 \nonumber & 
 + \text{three-fermion terms} \, ,
 \end{align}
while the corresponding transformation rules for the bosonic sector of the complex linear multiplet are 
\begin{align}
\label{dA} \delta A = & -  \sqrt{2} \left( \zeta \lambda + \overline\zeta \overline\rho \right)   \, , \\ 
\label{dF} \delta F = & -\frac{\sqrt{2}}{3} \overline M \zeta \lambda - \frac{i}{\sqrt{2}} \overline\zeta \overline\sigma^m  \hat{D}_m \lambda   - \frac{1}{2} \overline\zeta \overline\chi - \frac{1}{2 \sqrt{2}} b_m \overline\zeta \overline\sigma^m \lambda \, , \\
\label{dP} \nonumber \delta P_{\alpha \dot{\beta}} =& \, 2i \sqrt{2} (\sigma^m)_{\alpha \dot{\beta}} \zeta \hat{D}_m \lambda + i \sqrt{2} \, \zeta_\alpha \left[ (\hat{D}_m \lambda) \sigma^m \right]_{\dot{\beta}} + \zeta_\alpha \overline\chi_{\dot{\beta}} + \frac{\sqrt{2}}{6} b_m \left(\zeta \sigma^m \right)_{\dot{\beta}} \lambda_\alpha \\
& - \frac{1}{\sqrt{2}} \left( \sigma^m \right)_{\alpha \dot{\beta}} b_m \zeta \lambda - \frac{\sqrt{2}}{3} b_m \zeta_\alpha \left( \lambda \sigma^m \right)_{\dot{\beta}} +2i \sqrt{2} \,  \overline{\zeta}_{\dot{\beta}} ( \sigma^m \hat{D}_m \overline\rho )_\alpha \\
\nonumber & - \frac{4 \sqrt{2}}{3} M\overline{\zeta}_{\dot{\beta}}  \lambda_\alpha - \frac{4}{3} A \overline{\zeta}_{\dot{\beta}} \left( \sigma^{mn} \psi_{mn} \right)_\alpha - \frac{2i}{3} A b^m \overline{\zeta}_{\dot{\beta}} \psi_{m \alpha} + \frac{\sqrt{2}}{3} b_m \overline{\zeta}_{\dot{\beta}} \left(\sigma^m \overline\rho  
\right)_\alpha \, .
\end{align}
The derivative $\mathcal{D}_m$ is covariant with respect to local Lorentz transformations and we have defined 
\begin{align}
\hat{D}_m A \equiv & \, \partial_m A - \frac{1}{\sqrt{2}} \left( \psi_m \lambda +\overline\psi_m \overline\rho \right)    \, , \\
\hat{D}_m \lambda_\alpha \equiv & \, \mathcal{D}_m \lambda_\alpha - \frac{1}{\sqrt{2}} \psi_{m \alpha} F + \frac{1}{2 \sqrt{2}} \left( \sigma^n \overline\psi_m \right)_\alpha P_n \, , \\ 
\hat{D}_m \rho_\alpha \equiv & \, \mathcal{D}_m \rho_\alpha - \frac{i}{\sqrt{2}} \left( \sigma^n \overline\psi_m \right)_\alpha \hat{D}_n \overline A + \frac{1}{2 \sqrt{2}} \left(\sigma^n \overline\psi_m \right)_\alpha {\overline P}_n \, , \\
\nonumber \hat{D}_m P_{\alpha \dot{\beta}} \equiv & \, \mathcal{D}_m P_{\alpha \dot{\beta}} + i \sqrt{2} (\sigma^n)_{\alpha \dot{\beta}} \psi_m \hat{D}_n \lambda + \frac{i}{\sqrt{2}}  \psi_{m \alpha}
\left[ (\hat{D}_n \lambda) \sigma^n \right]_{\dot{\beta}} \\
&  + \frac{1}{2} \psi_{m \alpha} \overline\chi_{\dot{\beta}} + i \sqrt{2} \overline\psi_{m \dot{\beta}} ( \sigma^n\hat{D}_n \overline\rho )_\alpha \, , \\
\hat{D}_m \hat{D}_a A \equiv & \, \mathcal{D}_m \hat{D}_a A - \frac{1}{\sqrt{2}} \psi_m  \hat{D}_a \lambda - \frac{1}{\sqrt{2}} \overline\psi_m \hat{D}_a \overline\rho \, , \\
\hat{D}_m F \equiv & \, \partial_m F - \frac{i}{2 \sqrt{2}} \overline\psi_m \overline \sigma^n \hat{D}_n \lambda - \frac{1}{4} \overline\psi_m \overline \chi \, .
\end{align}
The most general kinetic Lagrangian for the complex linear superfield 
is given by  
\be
\label{Lkin} 
{\cal L}_\Omega= -\int d^4\theta\, E\ \Omega(\S , \overline{\S})\, , 
\ee 
where $\Omega(\Sigma, \overline \Sigma)$ is a real function of $\Sigma$ and $\overline\Sigma$. Making use of the superspace identity 
\be
\label{supidentity}
\int d^4\theta\, E \, Q = - \frac{1}{4}  \int d^2 \Theta \, 2 {\cal E} (\overline{\cal D}^2 -8 {\cal R}) Q \ + \  \text{total derivatives}\,, 
\ee
for any real superfield $Q$, 
we find that the bosonic sector of \eqref{Lkin} reads 
\begin{align}
\label{Lkincomp} \nonumber
e^{-1} \mathcal{L}^{(b)}_{\Omega} =& \, \frac{1}{6} \left( A \, \Omega_A + \overline A \, \Omega_{\overline A} - \Omega \right ) R - \Omega_{AA} \partial_m A \partial^m A - \Omega_{A \overline A}  \partial_m A \partial^m \overline A - \Omega_{\overline A \overline A} \partial_m \overline A \partial^m \overline A \\
\nonumber & + i \left( \Omega_{AA} P^m \partial_m A - \Omega_{\overline A \overline A} {\overline P}^m \partial_m \overline A \right) + \frac{1}{4} \left( \Omega_{AA}  P_m P^m + 2 \, \Omega_{A \overline A} P_m {\overline P}^m +\Omega_{\overline A \overline A}  {\overline P}_m  {\overline P}^m \right) \\
& - \Omega_{A \overline A} F \overline F + \frac{i}{3} b^m \left[ \Omega_{A \overline A} \left( \overline A \partial_m A - A \partial_m \overline A \right) - A \, \Omega_{AA} \partial_m A  + \overline A \, \Omega_{\overline A \overline A} \partial_m \overline A \right] \\ \nonumber
& - \frac{1}{9} \left( A \, \Omega_A + \overline A \, \Omega_{\overline A} - \Omega \right) b_m b^m - \frac{1}{3} \left( A M F \, \Omega_{AA} + \overline A \overline M \overline F \, \Omega_{\overline A \overline A} \right) \\
\nonumber & + \frac{1}{9} \left(A \, \Omega_A + \overline A \, \Omega_{\overline A} - \Omega - A  \overline A  \, \Omega_{A \overline A} \right) M \overline M   ,
\end{align}
where $\Omega=\Omega(A, \overline A)=\Omega(\Sigma, \overline \Sigma)|$, $\Omega_A = \frac{\partial \Omega}{\partial A}$,  $\Omega_{\overline A} = \frac{\partial \Omega}{\partial \overline A}$ etc. 

For the gravitational sector we have the typical superspace Lagrangian 
\be
\label{Langsugra}
{\cal L}_\text{SG} =  \int d^2 \Theta \, 2 {\cal E} \left( -3 {\cal R} + {\cal C} \right) + c.c.  \,, 
\ee
where we set $M_P=1$. 
The  chiral superfield ${\cal R}$ contains the Ricci scalar $R$, the supergravity auxiliary fields $M$ and $b_m$ and the gravitino $\psi_{m \alpha}$. 
With $2 {\cal E}$ we denote the chiral density in the conventions of  \cite{Wess:1992cp} and ${\cal C}$ is a real constant. 
The component form of the Lagrangian  \eqref{Langsugra} reads 
\be 
\begin{aligned}
e^{-1} {\cal L}_\text{SG} =&  - \frac{1}{2} R - \frac{1}{3} M \overline M + \frac{1}{3} b^m b_m - {\cal C} \left(  M + \overline M \right) \\
& + \frac{1}{4} \epsilon^{klmn} \left( \overline{\psi}_k \overline\sigma_l \psi_{mn} - \psi_k \sigma_l \overline{\psi}_{mn} \right) - \mathcal{C} \left( \psi_m \sigma^{mn} \psi_n + \overline\psi_m \overline\sigma^{mn}  \overline\psi_n \right) .
\end{aligned}
\ee

We now consider the theory described by the Lagrangian
\begin{equation}
\label{Lfree}
    \mathcal{L} = \mathcal{L}_\Omega + \mathcal{L}_{\text{SG}}\,, 
\end{equation}
and we want to find the on-shell  theory focusing first on the purely bosonic sector. 
As a first step we proceed to solve the equations of motion for the auxiliary field sector at zeroth order in the fermions.
The equation of motion for $M$ gives
\begin{equation}
\label{Meq}
M= - \, (3 \, \mathcal{C} + \overline F \, \overline A \, \Omega_{\overline A \overline A}) \, \mathcal{U} \, , 
\end{equation}
where 
\begin{equation}
\label{U}
\mathcal{U} \equiv \left[ 1 - \frac{1}{3} \left( A \, \Omega_A + \overline A \, \Omega_{\overline A} - \Omega - A \overline A \, \Omega_{A \overline A} \right) \right]^{-1} \,.
\end{equation} 
After substituting \eqref{Meq} into \eqref{Lfree}, the terms of the latter that contain $F$ or $\overline F$ are 
\begin{equation}
\label{LF}
e^{-1} \mathcal{L} \supset \frac{1}{3} \,  \mathcal{U} \left( 3\, \mathcal{C} + F A \,  \Omega_{AA}\right) (3 \, \mathcal{C} + \overline F \overline A \, \Omega_{\overline A \overline A}) - \Omega_{A \overline A} F \overline F \, . 
\end{equation}
Thus, varying the action with respect to $\overline F$ we obtain the following equation of motion 
\begin{equation}
\label{Feq}
F \left( \Omega_{A \overline A} - \frac{1}{3}  A \overline A |\Omega_{AA}|^2 \mathcal{U} \right) = \mathcal{C}  \overline A \, \Omega_{\overline A \overline A} \,  \mathcal{U} \, . 
\end{equation}
Furthermore, the equation of motion for $b_m$ gives
\begin{equation}
\label{beq}
b_m = \frac{3i}{2} \left( A \, \Omega_A + \overline A \, \Omega_{\overline A} - \Omega -3  \right)^{-1} \left[ \left( \overline A \, \Omega_{A \overline A} - A \, \Omega_{AA} \right) \partial_m A - \left( A \, \Omega_{A \overline A} - \overline A \, \Omega_{\overline A \overline A} \right) \partial_m \overline A \right] \,. 
\end{equation}
The equation of motion for $P_m$ reads 
\begin{equation}
\label{Peq}
P_m = -\frac{\Omega_{\overline A \overline A}}{\Omega_{A \overline A}} \left( {\overline P}_m - 2i \partial_m \overline A \right) \,, 
\end{equation} 
and is solved by 
\begin{equation}
\label{Ponshell}
P_m = 2 i \frac{\Omega_{\overline A \overline A}}{\Omega_{A \overline A}} \left( 1 - \frac{|\Omega_{AA}|^2}{\Omega_{A \overline A}^2} \right)^{-1} \left( \frac{\Omega_{A A}}{\Omega_{A \overline A}} \partial_m A + \partial_m \overline A \right) \, , 
\end{equation}
provided that\footnote{This condition for the case of $n$ complex linear multiplets is written as $(I=(i,\overline{i}),~J=(j,\overline{j}))$ 
\be
\det\Big(\Omega_{IJ}\Big)\neq 0,~~~~~~  I,J=1,\cdots, 2n.
\ee
}
\begin{equation}
\label{Pcondition}
 \Omega_{A \overline A}^2 \ne |\Omega_{AA}|^2 \,.   
\end{equation}
The theory \eqref{Lfree} is written in a Jordan frame, since 
\begin{equation}
e^{-1} \mathcal{L} \supset \frac{1}{6} \left( A \, \Omega_A + \overline A \, \Omega_{\overline A} - \Omega -3   \right) R \, , 
\end{equation}
so we need to perform the Weyl rescaling 
\begin{equation}
\label{Weylrescale}
{e_m}^a \ \rightarrow \ {\mathcal{S}}^{-\frac{1}{2}}   {e_m}^a \,, 
\end{equation}
where 
\begin{equation}
\label{S}
\mathcal{S} \equiv 1 - \frac{1}{3} \left( A \, \Omega_A + \overline A \, \Omega_{\overline A} - \Omega \right),
\end{equation}
to recover the correct normalisation for the Einstein--Hilbert term.

After integrating out the bosonic auxiliary fields and redefining the vielbein as in \eqref{Weylrescale},
the bosonic sector of \eqref{Lfree}
reads 
\be
\label{L(b)onshell}
\begin{aligned}
e^{-1} \mathcal{L}^{(b)} = & - \frac{1}{2} R - {\mathcal{S}}^{-1} \left[  \left( 1 - \frac{|\Omega_{AA}|^2}{\Omega_{A \overline A}^2} \right)^{-1} \Omega_{AA} + \frac{1}{3} {\mathcal{S}}^{-1} A \overline A \, \Omega_{AA} \Omega_{A \overline A} \right] \partial_m A \partial^m A \\
& - {\mathcal{S}}^{-1} \left[ \Omega_{A \overline A} + 2 \left( 1 - \frac{|\Omega_{AA}|^2}{\Omega_{A \overline A}^2} \right)^{-1} \frac{|\Omega_{AA}|^2}{\Omega_{A \overline A}} + \frac{1}{3}  {\mathcal{S}}^{-1} A \overline A \left( |\Omega_{AA}|^2 + \Omega_{A \overline A}^2 \right) \right] \partial_m A \partial^m \overline A \\
& -  {\mathcal{S}}^{-1}  \left[  \left( 1 - \frac{|\Omega_{AA}|^2}{\Omega_{A \overline A}^2} \right)^{-1} \Omega_{\overline A \overline A} + \frac{1}{3} {\mathcal{S}}^{-1} A \overline A \,  \Omega_{A \overline A} \Omega_{\overline A \overline A} \right] \partial_m \overline A \partial^m \overline A \\
& + \mathcal{C}^2 \mathcal{S}^{-2} \mathcal{U} \left[ 3 + A \overline A \, \mathcal{U}  |\Omega_{AA}|^2 \left( \Omega_{A \overline A} - \frac{1}{3} A \overline A \, \mathcal{U} |\Omega_{AA}|^2 \right)^{-1}\right] \,,
\end{aligned}
\ee
from where we read off the scalar potential 
\begin{equation}
\label{PotOmega}    
\mathcal{V} (A, \overline A)  = - \, \mathcal{C}^2 \mathcal{S}^{-2} \mathcal{U} \left[ 3 + A \overline A \, \mathcal{U}  |\Omega_{AA}|^2 \left( \Omega_{A \overline A} - \frac{1}{3} A \overline A \, \mathcal{U} |\Omega_{AA}|^2 \right)^{-1}\right] \,.
\end{equation}

To see the relation of \eqref{PotOmega} to the potential due to the coupling of 
standard scalar multiplets to supergravity we may consider the superspace duality Lagrangian 
\begin{equation}
\label{L'dual}
\mathcal{L}_{\text{dual}} = - \int d^4 \theta \, E \, \Omega ( \Sigma , \overline \Sigma   ) + \int d^4\theta\, E \left( \S \, \Phi + \overline \S \, \overline \Phi \right) + \left[  \int d^2 \Theta \, 2 {\cal E} \left( -3 {\cal R} + {\cal C} \right) + c.c. \right], 
\end{equation}
where $\Phi$ is a chiral superfield and $\Sigma$ is now unconstrained. If we vary \eqref{L'dual} with respect to $\Phi$, we obtain the linearity constraint \eqref{complin} for $\Sigma$ and we recover the theory \eqref{Lfree}. On the other hand, varying \eqref{L'dual} with respect to $\Sigma$ gives the equation of motion 
\begin{equation}
\frac{\partial \Omega}{\partial \Sigma} \left( \Sigma, \overline\Sigma\right) = \Phi\, ,    
\end{equation}
 which in principle can be solved for $\Sigma$ as a function of $\Phi$ and $\overline\Phi$, 
 \begin{equation}
 \label{S(Phi)}
\Sigma=\Sigma(\Phi, \overline \Phi)\, ,     
 \end{equation}
as has been pointed out in \cite{Deo:1985ix}. If we substitute \eqref{S(Phi)} into \eqref{L'dual}, the latter becomes
\begin{equation}
\label{L'Phi}
    \mathcal{L}_{\Phi}= \int d^4 \theta \,  E \, \tilde{\Omega} ( \Phi , \overline \Phi ) + \left[  \int d^2 \Theta \, 2 {\cal E} \left( -3 {\cal R} + {\cal C} \right) + c.c. \right],
\end{equation}
where 
\begin{equation}
     \tilde{\Omega} ( \Phi , \overline \Phi ) = \Phi \Sigma(\Phi, \overline \Phi) + \overline\Phi \overline\Sigma(\Phi, \overline \Phi) - \Omega(\Sigma(\Phi, \overline \Phi), \overline\Sigma(\Phi, \overline \Phi) ) \, .
\end{equation}
We conclude that \eqref{Lfree} is on-shell equivalent with \eqref{L'Phi} and therefore describes a standard supergravity chiral model.

In the following sections we will be primarily interested in the simple case where 
\begin{equation}
\label{SbarS}
\Omega(\Sigma, \overline \Sigma) = \Sigma \overline\Sigma \, . 
\end{equation} 
In this case, 
the expression for the kinetic Lagrangian \eqref{Lkin}, 
which we denote by $\mathcal{L}_{\Sigma \overline{\Sigma}}$, 
in terms of the component fields of the complex linear and the supergravity multiplets reads
\begin{align}
\label{Lang2}
\nonumber
e^{-1} {\cal L}_{\S \overline\Sigma} =& \frac{1}{6} A \overline A \, R + \frac{1}{2} P_m {\overline P}^m - F \overline F - \partial_m A \partial^m \overline A +\frac{i}{3}   b^m \left( \overline A \p_m A - A \p_m \overline A \right)    - \frac{1}{9} A \overline A \, b^m b_m \\ \nonumber
& - \frac{i}{2} \left(\rho \sigma^m \mathcal{D}_m \overline{\rho} + \overline{\rho} {\overline{\sigma}}^m \mathcal{D}_m \rho \right)  + \frac{1}{12} A \overline{A} \epsilon^{klmn} \left(\psi_k \sigma_l \overline{\psi}_{mn} - \overline{\psi}_k \overline{\sigma}_l \psi_{mn}\right) \\ \nonumber
& + \frac{\sqrt{2}}{3} \left( A \rho \sigma^{mn} \psi_{mn} + \overline{A} \overline{\rho} {\overline{\sigma}}^{mn} \overline{\psi}_{mn}\right) -\frac{\sqrt{2}}{2} \left( \psi_n \sigma^m \overline{\sigma}^n \rho \, \partial_m A + {\overline{\psi}}_n \overline{\sigma}^m \sigma^n \overline{\rho} \, \partial_m \overline{A}\right) \\
& + \frac{1}{4} \epsilon^{klmn} \left[ A \left(\partial_k \overline{A}\right) - \overline{A} \left( \partial_k A\right) \right] \psi_l \sigma_m {\overline{\psi}}_n + \frac{i\sqrt{2}}{6} b^m \left( A \, \psi_m \rho - \overline{A} \,  \overline{\psi}_m \overline{\rho} \right) \\ \nonumber
& - \frac{2}{3} b_m \, \lambda \sigma^m \overline{\lambda} - \frac{1}{6} b_m \, \rho \sigma^m \overline{\rho}  + \frac{i \sqrt{2}}{4} \left( \overline{F}  \lambda \sigma^m \overline{\psi}_m - F \psi_m \sigma^m \overline{\lambda} \right) \\ \nonumber
&+ \frac{3i \sqrt{2}}{8} \left( P^m \, \overline{\psi}_m \overline{\lambda} - \overline{P}^m \, \psi_m \lambda \right) + \frac{i\sqrt{2}}{4} \left( \psi_m \sigma^{mn} \lambda \, \overline{P}_n - \overline{\psi}_m \overline{\sigma}^{mn} \overline{\lambda} \, P_n \right) \\ \nonumber
&+ \frac{\sqrt{2}}{4} \left( \chi \lambda + \overline{\chi} \overline{\lambda} \right) + \text{four-fermion terms} \, .
\end{align}

We should also examine the case where the condition \eqref{Pcondition} does not hold. 
This happens if $\Omega(\Sigma, \overline\Sigma)=f(\Sigma+\overline\Sigma)$ for some real function $f$. 
We can illustrate this case with a simple example and therefore we take 
\begin{equation}
\label{S^2}
  \Omega(\Sigma, \overline\Sigma) =   (\Sigma + \overline \Sigma)^2 \,. 
\end{equation}
The bosonic sector of \eqref{Lfree} then reads
\be
\label{L'b_sing}
\begin{aligned}
 e^{-1} \mathcal{L}^{(b)} =& \frac{1}{6} \left[ \left(A + \overline A\right)^2 - 3 \right] R - 2 (\partial_m A \partial^m A + \partial_m A \partial^m \overline A + \partial_m \overline A \partial^m \overline A ) \\
& + 2i (P^m \partial_m A - \overline{P}^m \partial_m \overline A) + \frac{1}{2} \left( P_m P^m + 2 P_m {\overline P}^m + {\overline P}_m {\overline P}^m \right) \\
& + \frac{2i}{3} b^m \left( \overline A  - A   \right)\partial_m\left( A + \overline A \right) - \frac{1}{9} \left[ \left(A + \overline A\right)^2 - 3 \right] b_m b^m\\
& - 2 F \overline F - \frac{2}{3} \left( A M F + \overline A \overline M \overline F \right) + \frac{1}{9} \left(A^2 +{\overline A}^2 -3 \right) M \overline M - \mathcal{C} \left( M + \overline M \right) \,. 
\end{aligned}
\ee
After integrating out the scalar auxiliaries $F$ and $M$, \eqref{L'b_sing} becomes 
\begin{align}
\label{MandFintout} 
\nonumber e^{-1} \mathcal{L}^{(b)} =& \frac{1}{6} \left[ \left(A + \overline A\right)^2 - 3 \right] R - 2 (\partial_m A \partial^m A + \partial_m A \partial^m \overline A + \partial_m \overline A \partial^m \overline A ) \\
& + 2i (P^m \partial_m A - \overline{P}^m \partial_m \overline A) + \frac{1}{2} \left( P_m P^m + 2 P_m {\overline P}^m + {\overline P}_m {\overline P}^m \right) \\ \nonumber
& + \frac{2i}{3} b^m \left( \overline A  - A   \right) \partial_m\left( A + \overline A \right) - \frac{1}{9} \left[ \left(A + \overline A\right)^2 - 3 \right] b_m b^m \\ \nonumber
& - 9 \, \mathcal{C}^2 \left[ \left(A + \overline A\right)^2 - 3 \right]^{-1}.
 \end{align} 
 We then vary the action with respect to $P_m$ and we get the equation of motion
\begin{equation}
\label{P'eom}
P_m + {\overline P}_m = -2i \partial_m A  \,  , 
\end{equation}
which implies 
\begin{equation}
\label{ReA=const.}
A + \overline A = c \, ,
\end{equation}
where $c$ is a real constant. 
Substituting \eqref{ReA=const.} into \eqref{MandFintout} and using \eqref{P'eom} we find that all terms involving derivatives of the scalar $A$ cancel, so the Lagrangian is significantly simplified to
\begin{equation}
\label{MFPintout}
 e^{-1} \mathcal{L}^{(b)} =   \frac{1}{6} \left( c^2 - 3 \right) R  - \frac{1}{9} \left( c^2 - 3 \right) b_m b^m - 9 \, \mathcal{C}^2 \left( c^2 - 3 \right)^{-1} \, .
\end{equation}
In order for the graviton kinetic term to have the correct sign, the constant $c$ must satisfy
\begin{equation}
\label{c^2<3}
c^2<3 \,. 
\end{equation}
Furthermore, the equation of motion for $b_m$ has the trivial solution 
\begin{equation}
b_m=0 \, ,     
\end{equation}
which leads to 
\begin{equation}
\label{L'onshell}
 e^{-1} \mathcal{L}^{(b)} = \frac{1}{6} \left( c^2 - 3 \right) R  - 9 \, \mathcal{C}^2 \left( c^2 - 3 \right)^{-1} \, .
\end{equation}
After performing the Weyl rescaling
\begin{equation}
\label{Wresc}
{e_m}^a \ \rightarrow \  \left( 1 - \frac{c^2}{3}  \right)^{-\frac{1}{2}}   {e_m}^a \,, 
\end{equation}
which restores the proper normalisation for the graviton kinetic term, \eqref{L'onshell} becomes 
\begin{equation}
\label{L'resc}
e^{-1} \mathcal{L}^{(b)} = -\frac{1}{2} R + 3 \, \mathcal{C}^2  \left(1- \frac{c^2}{3}  \right)^{-3}.
\end{equation}
Hence, if  $\Omega(\Sigma, \overline \Sigma)$ is given by \eqref{S^2}, the bosonic sector of \eqref{Lfree} corresponds to pure anti-de Sitter gravity. 
Actually, in this case, 
none of the component fields of the complex linear multiplet $\Sigma$ is propagating and the theory \eqref{Lfree} reduces to pure AdS supergravity. 
To verify this we turn to superspace and we consider the duality Lagrangian 
\begin{equation}
\label{L''dual}
\mathcal{L}_{\text{dual}} =   - \int d^4 \theta E \, (\Sigma + \overline\Sigma )^2 + \int d^4\theta\, E \left( \S \, \Phi + \overline \S \, \overline \Phi \right) + \left[  \int d^2 \Theta \, 2 {\cal E} \left( -3 {\cal R} + {\cal C} \right) + c.c. \right], 
\end{equation}
where $\Phi$ is a chiral superfield and $\Sigma$ is now unconstrained. 
Varying \eqref{L''dual} with respect to $\Sigma$ we obtain 
\begin{equation}
\label{Phi=2c}
\Sigma+\overline\Sigma  = \frac{1}{2} \Phi = c \, , 
\end{equation}
where $c$ is a real constant, the same as the one appearing in \eqref{ReA=const.}. 
If we substitute \eqref{Phi=2c} into \eqref{L''dual}, the latter becomes
\begin{equation}
\label{Lsing}
\mathcal{L} = - (3-c^2) \int d^4 \theta E + \left(\int d^2 \Theta \, 2 {\cal E} {\cal C} + c.c.  \right),
\end{equation}
where we have also used \eqref{supidentity}. We can bring \eqref{Lsing} to the Einstein frame by redefining the super-determinant of the super-vielbein $E$ and the chiral density $2 \mathcal{E}$ as
\begin{equation}
\label{sdetredef}
E \rightarrow \left( 1 - \frac{c^2}{3} \right)^{-1} E \ , \quad 2 \mathcal{E} \rightarrow \left( 1 - \frac{c^2}{3} \right)^{-\frac{3}{2}}  2 \mathcal{E} , 
\end{equation}
which corresponds to a super-Weyl transformation with chiral superfield parameter $Y  = - \frac{1}{4} \ln \left( 1 - \frac{c^2}{3} \right)$ \cite{Howe}. Clearly $c$ must satisfy \eqref{c^2<3}. After the redefinitions \eqref{sdetredef} the Lagrangian \eqref{Lsing} becomes 
\begin{equation}
\mathcal{L} = -3 \int d^4 \theta E  +    \left[\int d^2 \Theta \, 2 {\cal E} \left( 1 - \frac{c^2}{3} \right)^{-\frac{3}{2}} {\cal C} + c.c.  \right] \, , 
\end{equation}
which corresponds to pure AdS$_4$supergravity.

\section{Supersymmetry breaking by the complex linear multiplet}

We now add to \eqref{Lfree} the following superspace higher derivative term  
\be
\label{ep1}
{\cal L}_\text{HD} = \int d^4\theta \, E \ \frac{1}{64 f^2 }\, 
{\cal D}^{\alpha}\S {\cal D}_{\alpha}\S \overline{{\cal D}}_{\dot{\alpha}}\overline{\S}\overline{{\cal D}}^{\dot{\alpha}}\overline{\S} \, , 
\ee
where for simplicity $f$ is a real constant. 
We could have a field-dependent $f$ but most parts of our analysis would remain unaltered, therefore we will not consider this possibility explicitly. 
The most important impact of a field-dependent $f$ is on the resultant scalar potential which would have an extra 
dependence on the scalars due to $f$. 
The term \eqref{ep1} is the supergravity counterpart of the higher derivative term for a complex linear superfield 
in global supersymmetry that was introduced in \cite{Farakos:2013zsa}. 
The model we will study reads
\begin{equation}
\label{Omega+SG+HD} 
\mathcal{L} = \mathcal{L}_\Omega + \mathcal{L}_{\text{SG}} + \mathcal{L}_{\text{HD}} \,. 
\end{equation}
The purely bosonic terms and the most important two-fermion terms of \eqref{ep1} are
\begin{align}
\label{ep3}
\nonumber e^{-1} {\cal L}_\text{HD} \supset &\frac{1}{64 f^2}\left( P_m P^m \overline{P}_n \overline{P}^n 
-8P_m\overline{P}^mF\overline{F}+16   F^2\overline{F}^2\right) \\ &-\frac{i}{8f^2} F \overline{F}\left(\lambda \sigma^m \mathcal{D}_m \overline{\lambda}+\overline{\lambda}\overline{\sigma}^m\mathcal{D}_m \lambda\right) - \frac{\sqrt{2}}{8f^2} F \overline{F} \left( \chi\lambda + \overline{\chi}\overline{\lambda} \right) \\\nonumber & + \frac{1}{12f^2} \left(\overline{M}\overline{F}^2\lambda\lambda + MF^2\overline{\lambda} \overline{\lambda} \right) \, .
\end{align} 
The full component form of \eqref{ep1} at second order in the fermions is given in the appendix. 
We see that even though we have explicitly introduced higher derivative terms there is no apparent Ostrogradsky instability
and the bosonic auxiliary fields will still have algebraic equations of motion. 
Notice that the structure of the term \eqref{ep1} allows it to be also written in the form 
\be
\label{ep1newFI}
{\cal L}_\text{HD} = \int d^4\theta \, E \ \frac{1}{64  f^2 }\, 
\frac{{\cal D}^{\alpha}\S {\cal D}_{\alpha}\S}{({\cal D}^2 \S)^2} 
\, 
\frac{\overline{{\cal D}}_{\dot{\alpha}}\overline{\S}\overline{{\cal D}}^{\dot{\alpha}}\overline{\S}}{(\overline {\cal D}^2 \overline \S)^2} 
\, ({\cal D}^2 \S)^2 (\overline {\cal D}^2 \overline \S)^2 \, , 
\ee 
which makes the similarity to \cite{Cribiori:2017laj} more apparent.

Our next task is to solve the equations of motion for the auxiliary field sector of the theory \eqref{Omega+SG+HD} ignoring terms with fermionic fields. The solutions of the equations of motion for the supergravity auxiliary fields $M$ and $b_m$ are still given by \eqref{Meq} and \eqref{beq} respectively, since the bosonic sector of the higher derivative term \eqref{ep1} does not involve these fields. After integrating out $M$ and $b_m$, the purely bosonic terms of \eqref{Omega+SG+HD} that contain at least one of the auxiliary fields $F$ and $P_m$ are
\be
\begin{aligned}
\label{LFP}
e^{-1} \mathcal{L} \supset & \frac{1}{3} \,  \mathcal{U} \left( 3\, \mathcal{C} + F A \,  \Omega_{AA}\right) (3 \, \mathcal{C} + \overline F \overline A \, \Omega_{\overline A \overline A}) - \Omega_{A \overline A} F \overline F
\\& + i \left( \Omega_{AA} P^m \partial_m A - \Omega_{\overline A \overline A} {\overline P}^m \partial_m \overline A \right) + \frac{1}{4} \left( \Omega_{AA}  P_m P^m + 2 \, \Omega_{A \overline A} P_m {\overline P}^m +\Omega_{\overline A \overline A}  {\overline P}_m  {\overline P}^m \right) \\
& + \frac{1}{64  f^2}\left( P_m P^m \overline{P}_n \overline{P}^n 
-8P_m\overline{P}^mF\overline{F}+16   F^2\overline{F}^2\right) \,,
\end{aligned}
\ee
where $\mathcal{U}$ is defined in \eqref{U}. The equations of motion for $F$ and $P_m$ thus read
\begin{equation}
\label{Feom}    
F \left( \Omega_{A \overline A} - \frac{1}{3}  A \overline A |\Omega_{AA}|^2 \mathcal{U} - \frac{1}{2f^2} F \overline F + \frac{1}{8f^2} P_m {\overline P}^m \right) = \mathcal{C}  \overline A \, \Omega_{\overline A \overline A} \,  \mathcal{U} \,, 
\end{equation}
and 
\begin{equation}
\label{Peom}    
\left( \Omega_{A \overline A} - \frac{1}{4f^2} F \overline F \right) P_m + \left(\Omega_{\overline A \overline A} + \frac{1}{16f^2} P_n P^n\right) {\overline P}_m = 2i \, \Omega_{\overline A \overline A} \partial_m \overline A \, , 
\end{equation}
respectively. 
Equations \eqref{Feom} and \eqref{Peom} are in general very difficult to solve. 
However, they are significantly simplified if the superspace kinetic function $\Omega(\Sigma,\overline\Sigma)$ is given by \eqref{SbarS}, 
in which case $\Omega_{A \overline A}=1$ and $\Omega_{AA}=\Omega_{\overline A \overline A}=0$, so \eqref{Peom} becomes 
\begin{equation}
\label{PeomSbarS}
\left( 1 - \frac{1}{4f^2} F \overline F  \right)P_m + \frac{1}{16f^2} P_n P^n {\overline P}_m = 0 \, . 
\end{equation}
The above equation is solved by 
\begin{equation}
\label{Pm=0}
P_m=0 \, , 
\end{equation}
while \eqref{Feom} reduces to
\be
\label{FeomSbarS}
F \left( 1 - \frac{1}{2 f^2} F \overline F \right) = 0 \, ,
\ee
after substituting \eqref{Pm=0}. 

From now on and up to and including section 5, the theory under consideration will be 
\begin{equation}
\label{total-old}    
\mathcal{L} = \mathcal{L}_{\Sigma \overline\Sigma} + \mathcal{L}_{\text{SG}} + \mathcal{L}_{\text{HD}}\, , \end{equation}
where
\begin{equation}
\label{LSbarS}
\mathcal{L}_{\Sigma \overline\Sigma} = -  \int d^4 \theta \, E \, \Sigma \overline\Sigma \, ,   
\end{equation}
whose component form has been given in \eqref{Lang2}. Making use of \eqref{Meq} and \eqref{beq} with $\Omega = A \overline A$ we find that the solutions of the equations of motion for the auxiliary fields $M$ and $b_m$ that arise from the bosonic sector of \eqref{total-old} are given by 
\begin{equation}
M = - 3 \, \mathcal{C}  \,, 
\end{equation}
and 
\begin{equation}
 b_m = \displaystyle\frac{3i}{2(A \overline{A}-3)} \left( \overline A \partial_m A - A \partial_m \overline A \right) \,. 
\end{equation}
Furthermore, if we ignore fermions, 
the equation of motion for $P_m$ is given by \eqref{PeomSbarS} and solved by \eqref{Pm=0}, 
while the equation of motion for $F$ is \eqref{FeomSbarS}. The latter has two solutions: 
\\[0.05cm]
\be
\label{oldv}
&& \text{{
\rm The standard vacuum}}: \ F=0 \,. 
\\[0.5cm] 
\label{newv}
&& \text{{
\rm A new vacuum}}: \ F \overline F = 2 \, f^2 \, .
\ee
\\[0.05cm] 
Therefore, integrating out the scalar auxiliary field of the complex linear multiplet leads to two inequivalent vacua. 
This is a novel aspect of the model \eqref{total-old}, 
which was originally found in global supersymmetry \cite{Farakos:2013zsa}. 
The standard vacuum does not offer any new properties other than a 
theory which can be dualized to a chiral model \cite{Farakos:2015vba}, as we will show later.
In the new vacuum supersymmetry is spontaneously broken,  
since 
\be
|F| = \sqrt{2}|f| \ne 0 .  
\ee

We now focus on the solution of the equation of motion for $F$ that corresponds to the supersymmetry breaking vacuum. 
The solutions of the equations of motion for the bosonic auxiliary fields at second order in the fermions have the following form 
\begin{align}
 \label{Pquad} P_m& = P_m^{(2)} \, , \\ 
 \label{bquad} b_m& = \displaystyle\frac{3i}{2(A \overline{A}-3)} \left( \overline A \partial_m A - A \partial_m \overline A \right) + b_m^{(2)} \, , \\
 \label{Mquad} M& = - 3 \, \mathcal{C} + M^{(2)} \, , \\
 \label{Fquad} F& = - \sqrt{2} f + F^{(2)} \, ,
\end{align}
where the quantities $P_m^{(2)},  b_m^{(2)}, M^{(2)} \, \text{and} \, F^{(2)}$ are quadratic in the fermions and we have chosen the expectation value of $F$ in the new vacuum to take the real value $\langle F \rangle = - \sqrt{2} f$. Since we are only interested in the terms of \eqref{total-old} that are at most quadratic in the fermions, it turns out that we do not need to specify any of  $P_m^{(2)},  b_m^{(2)}, M^{(2)}, F^{(2)}$  
because once we substitute equations \eqref{Pquad}-\eqref{Fquad} into \eqref{total-old}, the terms that are linear in these quantities and contain no other fermions cancel. After integrating out the bosonic auxiliary fields, the Lagrangian \eqref{total-old} becomes
\begin{align}
\label{Lonshell} \nonumber
e^{-1} \mathcal{L} = & \frac{1}{6} \left[ A \overline A - 3 + \frac{\sqrt{2}}{4f} \left( \overline A \lambda \lambda + A \overline{\lambda} \overline{\lambda} \right) \right] R - \partial_m A \, \partial^m \overline{A} - f^2 + 3 \, \mathcal{C}^2 \\ \nonumber
& -\frac{i}{2} \left(\rho \sigma^m \mathcal{D}_m \overline{\rho} + \overline{\rho} {\overline{\sigma}}^m \mathcal{D}_m \rho \right) - \frac{i}{4} \left ( \lambda \sigma^m \mathcal{D}_m \overline{\lambda} + \overline{\lambda} {\overline{\sigma}}^m \mathcal{D}_m \lambda \right) \\ \nonumber
& - \frac{1}{12} \left( A \overline{A} -3 \right) \epsilon^{klmn} \left( \overline\psi_k \overline\sigma_l \psi_{mn} - \psi_k \sigma_l \overline\psi_{mn}  \right)
+ \frac{\sqrt{2}}{3} \left( A  \rho \sigma^{mn} \psi_{mn} + \overline{A}  \overline{\rho} {\overline{\sigma}}^{mn} \overline{\psi}_{mn}\right) \\ \nonumber
&-\frac{\sqrt{2}}{2} \left( \psi_n \sigma^m \overline{\sigma}^n \rho \, \partial_m A + {\overline{\psi}}_n \overline{\sigma}^m \sigma^n \overline{\rho} \, \partial_m \overline{A}\right)  + \frac{1}{4} \epsilon^{klmn} \left( A \partial_k \overline{A} - \overline{A}  \partial_k A \right) \psi_l \sigma_m {\overline{\psi}}_n \\
& - \frac{i}{2} f \left( \lambda \sigma^m \overline\psi_m - \psi_m \sigma^m \overline\lambda \right) - \frac{1}{2} \, \mathcal{C} \left( \lambda \lambda + \overline\lambda \overline\lambda\right) - \mathcal{C} \left ( \psi_m \sigma^{mn} \psi_n + \overline\psi_m \overline\sigma^{mn} \overline\psi_n \right) \\ \nonumber
& - \frac{1}{4} \left( A \overline A - 3 \right)^{-1} \left( \overline A \partial_m A - A \partial_m \overline A \right) \Big{[} \overline A \partial^m A - A \partial^m \overline A + i \rho \sigma^m \overline\rho + \frac{i}{2} \lambda \sigma^m \overline\lambda \\ \nonumber
& \hspace{0.5cm} + \sqrt{2} \left(A \psi^m \rho - \overline{A} \overline\psi^m \overline\rho \right) \Big{]} - \frac{\sqrt{2}}{4f} \left [ \left( \partial_m \overline A \right) \partial^m \left( \lambda \lambda
\right) + \left( \partial_m A\right) \partial^m \left( \overline \lambda \overline\lambda \right) \right]  \\ \nonumber
& + \frac{\sqrt{2}}{8 f} \left( A \overline A - 3 \right)^{-1} \left( \overline A \partial_m A - A \partial_m \overline A \right) \Big{[} A \partial^m \left( \overline\lambda \overline\lambda \right) - \overline A \partial^m \left( \lambda \lambda \right) - \overline\lambda \overline\lambda \partial^m A + \lambda \lambda \partial^m \overline A \\ \nonumber
& \hspace{0.5cm} + \frac{1}{2} \left( A \overline A - 3 \right)^{-1} \left( \overline A \partial^m A - A \partial^m \overline A \right) \left( \overline A \lambda \lambda + A \overline \lambda \overline\lambda \right) \Big{]} \\ \nonumber
& + \text{terms at least quartic in the fermions} \, .
\end{align}
We note that since \eqref{ep1} involves the following coupling for the fermionic field $\lambda_\alpha$
\begin{equation}
 \mathcal{L}_{\text{HD}}   \supset  -\frac{i}{8f^2} e F \overline{F}\left(\lambda \sigma^m \mathcal{D}_m \overline{\lambda}+\overline{\lambda}\overline{\sigma}^m\mathcal{D}_m \lambda\right) , 
\end{equation}
the solution of the equation of motion for $F$ with nonzero purely bosonic term
$$F=-\sqrt{2} f + \text{fermions},$$
gives rise to a kinetic term for $\lambda_\alpha$ with the correct sign 
\begin{equation}
   \mathcal{L}  \supset  - \frac{i}{4} e \left ( \lambda \sigma^m \mathcal{D}_m \overline{\lambda} + \overline{\lambda} {\overline{\sigma}}^m \mathcal{D}_m \lambda \right) . 
\end{equation}
Therefore, in the new vacuum the previously auxiliary fermion $\lambda_\alpha$ becomes propagating.
We also introduce the Weyl fermion $\xi_\alpha = \displaystyle\frac{1}{\sqrt{2}} \lambda_\alpha$ to obtain a fermionic kinetic term with the correct coefficient 
\begin{equation}
     \mathcal{L}  \supset   - \frac{i}{2} e \left ( \xi \sigma^m \mathcal{D}_m \overline{\xi} + \overline{\xi} {\overline{\sigma}}^m \mathcal{D}_m  \xi \right) .
\end{equation}
 We will later identify $\xi_\alpha$ with the goldstino of the broken supersymmetry. Furthermore, we should point out that owing to \eqref{newv}, the two-fermion terms of \eqref{ep1} that involve the bilinears $\chi \lambda$ and $\overline \chi \overline \lambda$, 
\begin{equation}
    \mathcal{L}_{\text{HD}} \supset - \frac{\sqrt{2}}{8f^2} e F \overline{F} \left( \chi\lambda + \overline{\chi}\overline{\lambda} \right)  , 
\end{equation}
cancel the corresponding terms of \eqref{LSbarS},
\begin{equation}
\mathcal{L}_{\Sigma \overline\Sigma} \supset \frac{\sqrt{2}}{4}  e   \left( \chi\lambda + \overline{\chi}\overline{\lambda} \right)  ,
\end{equation}
thus protecting the theory from unwanted, dangerous terms. 
Moreover, we have a mass term for the fermion $\xi_\alpha$  
\begin{equation}
\mathcal{L} \supset - \frac{1}{2} \mathcal{C} e \left( \lambda \lambda + \overline \lambda \overline \lambda \right)  = - \,  \mathcal{C} e \left( \xi \xi + \overline \xi \overline \xi \right), 
\end{equation}
which originates from the last line of \eqref{ep3}. 
In addition, the terms of \eqref{Lonshell} that involve a factor of $\displaystyle\frac{1}{f}$ can be eliminated by the following redefinition of the scalar field $A$ 
\begin{equation}
\label{Aredef}
A = \overline B - \frac{\sqrt{2}}{4f} \lambda \lambda \, ,     
\end{equation}
where $B$ is a complex scalar field. The kinetic term for the graviton will then read  
\begin{equation}
 \mathcal{L}  \supset \frac{1}{6} e \left( B \overline B - 3 \right) R    \, , 
\end{equation}
so we perform the following Weyl rescaling of the vielbein
\begin{equation}
\label{e_m^a resc}
    {e_m}^a \rightarrow {\left( 1 - \frac{B \overline B}{3} \right)}^{-\frac{1}{2}} {e_m}^a \, , 
\end{equation}
to restore the canonical normalisation for the Einstein-Hilbert Lagrangian. We also redefine the fermionic fields as 
\begin{equation}
\label{fermion resc}
\xi_\alpha \rightarrow {\left( 1 - \frac{B \overline B}{3}\right) }^{\frac{1}{4}} \xi_\alpha 
\, , \  \rho_\alpha \rightarrow {\left( 1 - \frac{B \overline B}{3}\right) }^{\frac{1}{4}} \rho_\alpha 
\, , \ \psi_{m \alpha} \rightarrow {\left( 1 - \frac{B \overline B}{3}\right) }^{-\frac{1}{4}} \psi_{m \alpha} \, , 
\end{equation}
in order to recover the proper normalisation for their kinetic terms and we then perform an additional shift of the gravitino 
\begin{equation}
\label{gravitino shift}
 \psi_{m \alpha} \rightarrow  \psi_{m \alpha} + \frac{i \sqrt{2}}{6} B  {\left( 1 - \frac{B \overline B}{3}\right) }^{-1} \left( \sigma_m \overline\rho  \right)_\alpha \, ,  
\end{equation}
to eliminate the gravitino-fermion kinetic mixing terms that involve the 
fermion bilinears $\rho \sigma^{mn} \psi_{mn}$ and $\overline\rho \overline\sigma^{mn} \overline\psi_{mn}$. 
After all these field redefinitions, the Lagrangian \eqref{Lonshell} reads
\begin{align}
\label{Lrescaled}
 \nonumber e^{-1} \mathcal{L} =& -\frac{1}{2} R(e) - \left( 1 - \frac{B \overline B}{3} \right)^{-2} \partial_m B \partial^m \overline B \\ 
 \nonumber &- \frac{i}{2} \left( 1 - \frac{B \overline B}{3} \right)^{-2} \left( \rho \sigma^m \mathcal{D}_m \overline{\rho} + \overline{\rho} {\overline{\sigma}}^m \mathcal{D}_m  \rho \right)
 - \frac{i}{2} \left( 1 - \frac{B \overline B}{3} \right)^{-1} \left( \xi \sigma^m \mathcal{D}_m \overline{\xi} + \overline{\xi} {\overline{\sigma}}^m \mathcal{D}_m  \xi \right) \\
 \nonumber & + \frac{1}{2} \epsilon^{klmn} \left[ \overline\psi_k \overline\sigma_l \Tilde{\mathcal{D}}_m \psi_n - \psi_k \sigma_l \Tilde{\mathcal{D}}_m \overline\psi_n  + \frac{1}{2} \left( 1 - \frac{B \overline B}{3} \right)^{-1}  \left( \overline B  \partial_k B - B  \partial_k \overline B  \right) \psi_l \sigma_m \overline\psi_n \right] \\
\nonumber & + \frac{i}{12} \left( 1 - \frac{B \overline B}{3} \right)^{-3} \left( \overline B  \partial_m B - B  \partial_m \overline B\right) \left[ \rho \sigma^m \overline\rho - \left( 1 - \frac{B \overline B}{3} \right)\xi \sigma^m \overline\xi  \right] \\
 & - \frac{\sqrt{2}}{2} \left( 1 - \frac{B \overline B}{3} \right)^{-2} \left ( \rho \sigma^m \overline \sigma^n \psi_m \, \partial_n  \overline B + \overline\rho \overline\sigma^m \sigma^n \overline \psi_m \, \partial_n B \right) \\
\nonumber & - \left( 1 - \frac{B \overline B}{3} \right)^{-\frac{3}{2}} \bigg[ \, \mathcal{C} \left( \psi_m \sigma^{mn} \psi_n + \overline\psi_m \overline\sigma^{mn} \overline\psi_n \right)   + \frac{i\sqrt{2}}{2} f \left( \xi \sigma^m \overline\psi_m + \overline\xi \overline\sigma^m \psi_m \right) \\
\nonumber & \hspace{0.5 cm} + \frac{i\sqrt{2}}{2} \, \mathcal{C} \left( 1 - \frac{B \overline B}{3} \right)^{-1} \left( \overline B \rho \sigma^m \overline\psi_m + B \overline\rho \overline\sigma^m \psi_m \right) + \mathcal{C} \left ( \xi \xi + \overline{\xi} \overline{\xi}  \right) \\
\nonumber & \hspace{0.5cm} + \frac{2}{3} f \left( 1 - \frac{B \overline B}{3} \right)^{-1} \left( \overline B \, \xi \rho + B \, \overline\xi \overline\rho \right)  + \frac{1}{3} \, \mathcal{C} \left( 1 - \frac{B \overline B}{3} \right)^{-2} \left( {\overline B}^2 \rho \rho + B^2 \overline \rho \overline \rho \right) \bigg] \\
\nonumber & \hspace{0.5cm} - \left( 1 - \frac{B \overline B}{3} \right)^{-2} \left( f^2 - 3 \, \mathcal{C}^2 \right)  + \text{terms at least quartic in the fermions} \, ,
 \end{align}
where  the curvature scalar $R(e)$ contains the torsion-free spin connection and we have used the fact that 
\begin{equation}
e R = e R(e) + \text{total derivative} + \text{four-fermion terms}    
\end{equation}
omitting the total derivative.
Therefore, the scalar potential in the Einstein frame is given by 
\begin{equation}
\label{Potential}
\mathcal{V} (B, \overline B) =  \left( f^2 - 3 \, \mathcal{C}^2 \right) \left( 1 - \frac{B \overline B}{3} \right)^{-2} \, ,
\end{equation}
which for 
\begin{equation}
    f^2 > 3 \, \mathcal{C}^2
\end{equation}
has a global minimum at $B=0$, which corresponds to a de Sitter vacuum, since 
\begin{equation}
    \label{vevV}
    \langle \mathcal{V} \rangle = \mathcal{V} (0,0) = f^2 - 3 \, \mathcal{C}^2 > 0 \, .
\end{equation}

Furthermore, after integrating out the auxiliary fields and performing the field redefinitions \eqref{Aredef} and \eqref{e_m^a resc}-\eqref{gravitino shift}, the supergravity transformations of the propagating spin-$\frac{1}{2}$ fermions $\xi_\alpha$ and $\rho_\alpha$ read 
\begin{align}
\label{xishift}
\delta \xi_\alpha &= \sqrt{2} f \left( 1 - \frac{B \overline B}{3} \right)^{-\frac{1}{2}} \zeta_\alpha + \dots \, , \\ 
\label{rhoshift}
\delta \rho_\alpha & = \sqrt{2} \, \mathcal{C} B \left( 1 - \frac{B \overline B}{3} \right)^{-\frac{1}{2}} \zeta_\alpha + \dots \, , 
\end{align}
where we have also redefined the transformation parameters as $\zeta_\alpha \rightarrow \left( 1 - \frac{B \overline B}{3} \right)^{-\frac{1}{4}}  \zeta_\alpha $ and the omitted terms involve derivatives of the scalar $B$ or are at least cubic in the fermions. Since $\langle B \rangle=0$, we have that in the vacuum
\begin{equation}
\label{shift vevs}
\delta \xi_\alpha  = \sqrt{2} f \zeta_\alpha \ne 0 \ , \quad  \delta \rho_\alpha  = 0 \, , 
\end{equation}
so the Goldstone mode of the spontaneously broken supersymmetry is the fermion $\xi_\alpha$, 
which undergoes a shift under local supersymmetry transformations. 
This goldstino is one of the auxiliary fields which is propagating only in the supersymmetry breaking vacuum 
and also has no superpartner,  so supersymmetry is intrinsically non-linearly realized as in \cite{Farakos:2013zsa}.

In order to deduce the mass of the complex scalar $B$ in the supersymmetry breaking vacuum, we Taylor expand the potential \eqref{Potential} around its minimum point, $(B, \overline B) = (0,0)$, 
\begin{equation}
\label{VTaylor}
\mathcal{V} (B, \overline B) =  \left( f^2 - 3 \, \mathcal{C}^2 \right) \left( 1 + \frac{2 B \overline B}{3} + \mathcal{O}((B \overline B)^2) \right)  .
\end{equation}
The term on the right hand side of \eqref{VTaylor}
that is quadratic in $B$ indicates that the squared mass of this field equals
\begin{equation}
\label{mB}
m_B^2 = \frac{2}{3} \left ( f^2 - 3 \, \mathcal{C}^2 \right)  . 
\end{equation}
Since $\langle B \rangle = 0$, the fermion mass terms in the vacuum \eqref{newv} read
\be
\label{fermionmass}
\begin{aligned}
e^{-1} \mathcal{L}_{\text{fermion masses}} =& - \, \mathcal{C} \left(\psi_m \sigma^{mn} \psi_n + \overline\psi_m \overline\sigma^{mn} \overline\psi_n \right) - \frac{i\sqrt{2}}{2} f \left( \xi \sigma^m \overline\psi_m + \overline\xi \overline\sigma^m \psi_m \right) \\
& - \, \mathcal{C} \left( \xi \xi + \overline\xi \overline\xi \right) , 
\end{aligned}
\ee
so there exists a mixing term between the gravitino and the goldstino $\xi_\alpha$. In order to ``diagonalize'' the Lagrangian, we follow a procedure similar to the one described in \cite{Ferrara:2016}. We perform the following redefinitions of the gravitino and the Weyl fermion $\rho_\alpha$
\begin{align}
\label{gravredef}
\psi_{m \alpha} \rightarrow \, & \psi_{m \alpha} - \frac{\sqrt{2}}{f} \mathcal{D}_m \left[ \left( 1 - \frac{B \overline B}{3} \right)^{\frac{1}{2}} \xi_\alpha \right] - \frac{\sqrt{2}}{4f} \left( 1 - \frac{B \overline B}{3} \right)^{- \frac{1}{2}} \left( \overline B \partial_m B - B \partial_m \overline B \right) \xi_\alpha \\ \nonumber & - \frac{i \,  \mathcal{C} }{\sqrt{2} f} \left( 1 - \frac{B \overline B}{3} \right)^{- 1} \left( \sigma_m \overline \xi \right)_\alpha, \\
\label{rhoredef} \rho_\alpha \rightarrow \, & \rho_\alpha + \frac{\mathcal{C}}{f} B \xi_\alpha - \frac{i}{f} \left( 1 - \frac{B \overline B}{3} \right)^{ \frac{1}{2}} \left( \sigma^m \overline \xi \right)_\alpha \partial_m B \, , 
\end{align}
followed by shifts of the vielbein and the scalar $B$ 
\begin{align}
\label{vielshift}
{e_m}^a \rightarrow \, & {e_m}^a - \frac{i}{\sqrt{2} f} \left( 1 - \frac{B \overline B}{3} \right)^{ \frac{1}{2}} \left( \xi \sigma^a \overline\psi_m + \overline \xi \overline\sigma^a \psi_m \right), \\
\label{Bshift}
B \rightarrow  \, & B - \frac{1}{f} \left( 1 - \frac{B \overline B}{3} \right)^{ \frac{1}{2}} \xi \rho \, .
\end{align}
The field redefinitions \eqref{gravredef}-\eqref{Bshift} correspond to a local supersymmetry transformation with parameters 
\begin{equation}
\zeta_\alpha = - \frac{1}{\sqrt{2} f} \left( 1 - \frac{B \overline B}{3} \right)^{ \frac{1}{2}} \xi_\alpha  \,  , 
\end{equation}
such that
\begin{equation}
\xi_\alpha + \delta_{\text{lin}} \xi_\alpha = 0 \, ,     
\end{equation}
where $\delta_{\text{lin}} \xi_\alpha$ is the part of the supergravity transformation of the goldstino $\xi_\alpha$ that is linear in the fermions written explicitly in \eqref{xishift}. 
Using the identity 
\begin{equation}
\label{[Dm,Dn]}
[\mathcal{D}_m, \mathcal{D}_n] \eta_\alpha = \frac{1}{2} R_{mnab}( \sigma^{ab} \eta )_\alpha \, ,
\end{equation} 
where $\eta_\alpha$ is an arbitrary Weyl spinor, 
one can show that after these redefinitions all the two-fermion couplings of the goldstino with the fermion $\rho_\alpha$ 
and with the gravitino are eliminated up to a total derivative.

After the previous field redefinitions that eliminated the gravitino-goldstino mixing, 
the resulting Lagrangian involves the following  quadratic in the goldstino terms  
\begin{align}
\label{Lxixi} \nonumber
 e^{-1} \mathcal{L} \supset  &  \left(  f^2 - 3 \, \mathcal{C}^2 \right) \bigg{[}\frac{\mathcal{C}}{f^2} \left( 1 - \frac{B \overline B}{3} \right)^{-\frac{3}{2}} \left( \xi \xi + \overline\xi \overline\xi \right) + \frac{i}{2 f^2} \left( 1 - \frac{B \overline B}{3} \right)^{-1} \left( \xi \sigma^m \mathcal{D}_m \overline\xi + \overline\xi \overline\sigma^m \mathcal{D}_m \xi \right) 
    \\ \nonumber & \hspace{0.2cm} + \frac{i}{12 f^2}\left( 1 - \frac{B \overline B}{3} \right)^{-2} \left( \overline B \partial_m B - B \partial_m \overline B \right) \xi \sigma^m \overline\xi \bigg{]} - \frac{\mathcal{C}}{f^2} \left( 1 - \frac{B \overline B}{3} \right)^{-1} \times \\ \nonumber & \Bigg{\{} 2 \, \xi \sigma^{mn} \mathcal{D}_m \mathcal{D}_n \left[\left( 1 - \frac{B \overline B}{3} \right)^{\frac{1}{2}} \xi \right] + \frac{1}{12} \left( 1 - \frac{B \overline B}{3} \right)^{-\frac{3}{2}} B \partial_m \overline B \left( \overline B \partial^m B - B \partial^m \overline B \right) \xi \xi \\ \nonumber
    & \hspace{0.2cm} + \left( 1 - \frac{B \overline B}{3} \right)^{-\frac{1}{2}} B \left( \partial^m \overline B \right) \xi \mathcal{D}_m \xi + c.c. \Bigg{\}} + \frac{1}{f^2} \epsilon^{klmn} \left( 1 - \frac{B \overline B}{3} \right)^{\frac{1}{2}} \times \\ 
   & \Bigg{\{}  \overline\xi \overline\sigma_k \mathcal{D}_l \mathcal{D}_m \mathcal{D}_n \left[\left( 1 - \frac{B \overline B}{3} \right)^{\frac{1}{2}} \xi \right] + c.c. \Bigg{\}} - \frac{1}{2 f^2} \epsilon^{klmn} \left( 1 - \frac{B \overline B}{3} \right)^{-\frac{1}{2}} \times \\ \nonumber
   & \left( \overline B \partial_k B - B \partial_k \overline B  \right) \xi \sigma_l \mathcal{D}_m \mathcal{D}_n \left[\left( 1 - \frac{B \overline B}{3} \right)^{\frac{1}{2}} \overline\xi \right] + \frac{i}{2 f^2} \left( 1 - \frac{B \overline B}{3} \right)^{-1} \times \\
   \nonumber & \bigg{\{} \partial_m B \partial^m \overline B ( \xi \sigma^n \mathcal{D}_n \overline\xi + \overline\xi \overline\sigma^n \mathcal{D}_n \xi ) - 2 \partial^m B \partial^n \overline B (\xi \sigma_{(m} \mathcal{D}_{n)} \overline \xi + \overline \xi \overline\sigma_{(m}\mathcal{D}_{n)} \xi ) \\ \nonumber & \hspace{0.12cm} + (\partial_n B \partial^m \overline B - \partial_n \overline B \partial^m B) \partial_m (\xi \sigma^n \overline \xi) + \left[(\partial_m \partial_n B) \partial^m \overline B - (\partial_m \partial_n \overline B) \partial^m B \right]\xi \sigma^n \overline \xi \bigg{\}} \\ \nonumber & + \frac{i}{f^2} \left( 1 - \frac{B \overline B}{3} \right)^{-2} \bigg{[} \frac{1}{6} \big{(} \overline B \partial_n B - B \partial_n \overline B \big{)} \partial_m B \partial^m \overline B + \frac{5}{12} \big{(} \overline B \partial_m B \partial^m B \partial_n \overline B \\ \nonumber
   & \hspace{0.2cm}- B \partial_m \overline B \partial^m \overline B \partial_n B \big{)}\bigg{]} \xi \sigma^n \overline \xi, 
  \end{align} 
which can be eliminated if we further redefine the bosonic fields as 
\begin{align}
\label{e2ndshift} \nonumber
{e_m}^a \rightarrow \, & {e_m}^a + \bigg{\{ } \frac{i}{2 f^2} \left( 1 - \frac{B \overline B}{3} \right)^{\frac{1}{2}}  \overline \xi \overline\sigma^a \mathcal{D}_m \bigg{[}\left( 1 - \frac{B \overline B}{3} \right)^{\frac{1}{2}} \xi \bigg{]} \\
& - \frac{i}{8 f^2} \left( \overline B \partial_m B - B \partial_m \overline B \right) \xi \sigma^a \overline\xi + \frac{\mathcal{C}}{4 f^2} \left( 1 - \frac{B \overline B}{3} \right)^{-\frac{1}{2}} \xi \xi {e_m}^a + c.c. \bigg{\}} \\
B \rightarrow \, & B - \frac{\mathcal{C}}{2 f^2} \left( 1 - \frac{B \overline B}{3} \right)^{\frac{1}{2}} B \xi \xi + \frac{i}{2 f^2} \left( 1 - \frac{B \overline B}{3} \right) \xi \sigma^m \overline\xi \partial_m B .
\end{align}
The above redefinitions give rise to additional terms quadratic in the goldstino,
\begin{align} 
\label{newxixi}
- & \frac{1}{2} e R(e) - e \left( 1 - \frac{B \overline B}{3} \right)^{-2} \partial_m B \partial^m \overline B - e \mathcal{V} \\
\ \nonumber
\rightarrow & - \frac{1}{2} e R(e) - e \left( 1 - \frac{B \overline B}{3} \right)^{-2} \partial_m B \partial^m \overline B - e \mathcal{V} \\
\ \nonumber
& -  \left(  f^2 - 3 \, \mathcal{C}^2 \right) e \bigg{[}\frac{\mathcal{C}}{f^2} \left( 1 - \frac{B \overline B}{3} \right)^{-\frac{3}{2}} \left( \xi \xi + \overline\xi \overline\xi \right) + \frac{i}{2 f^2} \left( 1 - \frac{B \overline B}{3} \right)^{-1} \left( \xi \sigma^m \mathcal{D}_m \overline\xi + \overline\xi \overline\sigma^m \mathcal{D}_m \xi \right) 
    \\
    \ \nonumber & \hspace{0.33cm} + \frac{i}{12 f^2}\left( 1 - \frac{B \overline B}{3} \right)^{-2} \left( \overline B \partial_m B - B \partial_m \overline B \right) \xi \sigma^m \overline\xi \bigg{]} + \frac{\mathcal{C}}{f^2} e \left( 1 - \frac{B \overline B}{3} \right)^{-\frac{1}{2}} \bigg{[} - \frac{1}{4} R(e) \xi \xi \\ \nonumber 
    & \hspace{0.33 cm} + \frac{1}{12} \left( 1 - \frac{B \overline B}{3} \right)^{-2} B \partial_m \overline B \left( \overline B \partial^m B - B \partial^m \overline B \right) \xi \xi + \left( 1 - \frac{B \overline B}{3} \right)^{-1} B \left(\partial^m \overline B \right) \xi \mathcal{D}_m \xi + c.c. \bigg{]} \\ \nonumber 
& + \frac{i}{2 f^2} e \Bigg{\{}  \left( 1 - \frac{B \overline B}{3} \right)^{\frac{1}{2}} \Bigg{\{} \xi \sigma^n \mathcal{D}_m \left[ \left(1-\frac{B \overline B}{3} \right)^{\frac{1}{2}} \overline \xi \right] +  \overline\xi \overline\sigma^n \mathcal{D}_m \left[ \left(1-\frac{B \overline B}{3} \right)^{\frac{1}{2}} \xi \right]\Bigg{\}} \\ \nonumber 
& \hspace{0.33cm} - \frac{1}{2} \left( \overline B \partial_m B - B \partial_m \overline B  \right) \xi \sigma^n \overline \xi \Bigg{\}} \left[ {R_n}^m (e) - \frac{1}{2} \delta_n^m R(e) \right] - \frac{i}{2 f^2} e \left( 1 - \frac{B \overline B}{3} \right)^{-1} \times \\ \nonumber 
   & \big{\{} \partial_m B \partial^m \overline B ( \xi \sigma^n \mathcal{D}_n \overline\xi + \overline\xi \overline\sigma^n \mathcal{D}_n \xi ) - 2 \partial^m B \partial^n \overline B (\xi \sigma_{(m} \mathcal{D}_{n)} \overline \xi + \overline \xi \overline\sigma_{(m}\mathcal{D}_{n)} \xi ) \\  \nonumber  & \hspace{0.12cm} + (\partial_n B \partial^m \overline B - \partial_n \overline B \partial^m B) \partial_m (\xi \sigma^n \overline \xi) + \left[(\partial_m \partial_n B) \partial^m \overline B - (\partial_m \partial_n \overline B) \partial^m B \right]\xi \sigma^n \overline \xi \big{\}} \\  \nonumber  & - \frac{i}{f^2} e \left( 1 - \frac{B \overline B}{3} \right)^{-2} \bigg{[} \frac{1}{6} \big{(} \overline B \partial_n B - B \partial_n \overline B \big{)} \partial_m B \partial^m \overline B + \frac{5}{12} \big{(} \overline B \partial_m B \partial^m B \partial_n \overline B \\ \nonumber 
   & \hspace{0.2cm}- B \partial_m \overline B \partial^m \overline B \partial_n B \big{)}\bigg{]}\xi \sigma^n \overline \xi, 
\end{align} 
where $R_{mn}(e)$ is the Ricci tensor associated with the torsion-free spin connection. Making use of \eqref{[Dm,Dn]} one can show that the last nine lines of \eqref{newxixi} cancel the terms \eqref{Lxixi} if we neglect terms of order greater than two in the fermions due to torsion. Thus, the Lagrangian \eqref{Lrescaled} eventually becomes 
\begin{align}
\label{xi=0 gauge} \nonumber
e^{-1} \mathcal{L} =  & -\frac{1}{2} R (e) - \left( 1 - \frac{B \overline B}{3} \right)^{-2} \partial_m B \partial^m \overline B \\ \nonumber
   &- \frac{i}{2} \left( 1 - \frac{B \overline B}{3} \right)^{-2} \left( \rho \sigma^m \mathcal{D}_m \overline{\rho} + \overline{\rho} {\overline{\sigma}}^m \mathcal{D}_m  \rho \right)
    + \frac{1}{2} \epsilon^{klmn} \left( \overline\psi_k \overline\sigma_l \Tilde{\mathcal{D}}_m \psi_n - \psi_k \sigma_l \Tilde{\mathcal{D}}_m \overline\psi_n  \right) \\ \nonumber
 & + \frac{1}{4} \epsilon^{klmn} \left( 1 - \frac{B \overline B}{3} \right)^{-1}  \left( \overline B  \partial_k B - B  \partial_k \overline B  \right) \psi_l \sigma_m \overline\psi_n \\  &  + \frac{i}{12} \left( 1 - \frac{B \overline B}{3} \right)^{-3} \left( \overline B  \partial_m B - B  \partial_m \overline B\right)  \rho \sigma^m \overline\rho   \\ \nonumber & - \frac{\sqrt{2}}{2} \left( 1 - \frac{B \overline B}{3} \right)^{-2} \left ( \rho \sigma^m \overline \sigma^n \psi_m \, \partial_n  \overline B + \overline\rho \overline\sigma^m \sigma^n \overline \psi_m \, \partial_n B \right) \\ \nonumber
 & -\mathcal{C} \left( 1 - \frac{B \overline B}{3} \right)^{-\frac{3}{2}}  \bigg[  \left( \psi_m \sigma^{mn} \psi_n + \overline\psi_m \overline\sigma^{mn} \overline\psi_n \right)   + \frac{i\sqrt{2}}{2}  \left( 1 - \frac{B \overline B}{3} \right)^{-1} \left( \overline B \rho \sigma^m \overline\psi_m + B \overline\rho \overline\sigma^m \psi_m \right)  \\ \nonumber
 & \hspace{0.5cm}  + \frac{1}{3} \, \left( 1 - \frac{B \overline B}{3} \right)^{-2} \left( {\overline B}^2 \rho \rho + B^2 \overline \rho \overline \rho \right) \bigg]  - \left( 1 - \frac{B \overline B}{3} \right)^{-2} \left( f^2 - 3 \, \mathcal{C}^2 \right) \\
 \nonumber & + \text{terms at least quartic in the fermions}.
\end{align}
We note that the goldstino no longer appears in any of the terms that are quadratic in the fermions. 
It is possible to eliminate the terms of higher order in the fermions involving the goldstino as well. 
This is the super-BEH mechanism, in which the goldstino is ``eaten'' by the gravitino. 
The physical field content of the theory in the vacuum \eqref{newv} consists of a massive scalar $B$, a massless Weyl fermion $\rho_\alpha$, a massive gravitino with mass $m_\psi = \mathcal{C}$ and a massless graviton.

We also note that when
\begin{equation}
\label{f^2=3C^2}
    f^2=3 \, \mathcal{C}^2 \, , 
\end{equation}
the scalar potential \eqref{Potential} vanishes identically, 
\begin{equation}
\mathcal{V}=0\,  ,   
\end{equation}
so the choice \eqref{f^2=3C^2} gives a new no-scale supergravity \cite{Cremmer:1983bf,Dall'Agata:2013ksa}. In these theories supersymmetry is spontaneously broken with a vanishing vacuum energy, but the supersymmetry breaking scale is not fixed at the tree-level, since the gravitino mass is also not fixed.

Finally, if $f^2<3 \, \mathcal{C}^2$, the scalar potential has a global maximum at $B=0$ corresponding to an anti-de Sitter vacuum, since $\langle \mathcal{V} \rangle= f^2- 3 \mathcal{C}^2 < 0 $, and the scalar field $B$ has negative squared mass given by \eqref{mB}, which satisfies the BF bound \cite{Breitenlohner:1982-197,Breitenlohner:1982-249}, since
\begin{equation}
\label{BFbound}
m_B^2 > \frac{3}{4} \langle \mathcal{V}   \rangle \, . 
\end{equation}
We thus have a perturbatively stable anti-de Sitter vacuum with broken supersymmetry.

\section{Analysis around the standard vacuum}

Let us now study the theory \eqref{total-old} around the standard vacuum, where $\langle F \rangle = 0$. At second order in the fermions  the corresponding solutions of the equations of motion for the bosonic auxiliary fields have the forms \eqref{Pquad}-\eqref{Mquad} and $F=F^{(2)}$, where $F^{(2)}$ is a quantity quadratic in the fermionic fields. Substituting these solutions into the Lagrangian \eqref{total-old} leads to 
\begin{align}
\label{Loldv}
 \nonumber
e^{-1} \mathcal{L} = & \frac{1}{6} \left( A \overline A - 3 \right) R - \partial_m A \, \partial^m \overline{A} + 3 \, \mathcal{C}^2 -\frac{i}{2} \left(\rho \sigma^m \mathcal{D}_m \overline{\rho} + \overline{\rho} {\overline{\sigma}}^m \mathcal{D}_m \rho \right)  \\ \nonumber
& - \frac{1}{12} \left( A \overline{A} -3 \right) \epsilon^{klmn} \left( \overline\psi_k \overline\sigma_l \psi_{mn} - \psi_k \sigma_l \overline\psi_{mn}  \right)
+ \frac{\sqrt{2}}{3} \left( A  \rho \sigma^{mn} \psi_{mn} + \overline{A}  \overline{\rho} {\overline{\sigma}}^{mn} \overline{\psi}_{mn}\right) \\ \nonumber
&-\frac{\sqrt{2}}{2} \left( \psi_n \sigma^m \overline{\sigma}^n \rho \, \partial_m A + {\overline{\psi}}_n \overline{\sigma}^m \sigma^n \overline{\rho} \, \partial_m \overline{A}\right)  + \frac{1}{4} \epsilon^{klmn} \left( A \partial_k \overline{A} - \overline{A}  \partial_k A \right) \psi_l \sigma_m {\overline{\psi}}_n  \\ 
& - \frac{1}{4} \left( A \overline A - 3 \right)^{-1} \left( \overline A \partial_m A - A \partial_m \overline A \right) \Big{[} \overline A \partial^m A - A \partial^m \overline A + i \rho \sigma^m \overline\rho  \\ \nonumber
& \hspace{0.5cm} + \sqrt{2} \left(A \psi^m \rho - \overline{A} \overline\psi^m \overline\rho \right) \Big{]} - \mathcal{C} \left ( \psi_m \sigma^{mn} \psi_n + \overline\psi_m \overline\sigma^{mn} \overline\psi_n \right) 
\\ \nonumber & - i \left( A \overline A - 3 \right)^{-1} \left( \overline A \partial_m A - A \partial_m \overline A \right) \lambda \sigma^m \overline \lambda + \frac{\sqrt{2}}{4} \left( \chi \lambda + \overline \chi \overline \lambda \right)
\\ \nonumber
& + \text{terms at least quartic in the fermions}\,.
\end{align}
We then integrate out the auxiliary fermions $\lambda_\alpha$ and $\chi_\alpha$ to obtain the following equations of motion
\begin{align}
\delta \lambda^\alpha &:\ \ \chi_\alpha = 2 i \sqrt{2}  \left( A \overline A - 3 \right)^{-1} \left( \overline A \partial_m A - A \partial_m \overline A \right) \left( \sigma^m \overline\lambda \right)_\alpha  + \dots   \\
\delta \chi^\alpha &: \ \ \lambda_\alpha=0 + \dots \, ,
\end{align}
where the dots represent terms that are at least cubic in the fermions. Therefore, on-shell we have that $\lambda_\alpha=\chi_\alpha=0$  at first order in the fermions. We will later show using superspace methods that to the standard vacuum there correspond the exact solutions $F=P_{\alpha \dot{\alpha}}=\lambda_\alpha=\chi_\alpha=0 $ of the equations of motion for the auxiliary field sector. 

In order to restore the correct normalisation for the kinetic terms of the propagating fields, we perform the Weyl rescalings
\begin{equation}
\label{Weylresc}
{e_m}^a \rightarrow \left( 1 - \frac{A \overline A}{3} \right)^{-\frac{1}{2}} {e_m}^a, ~~~~ \rho_\alpha \rightarrow \left( 1 - \frac{A \overline A}{3} \right)^{\frac{1}{4}} \rho_\alpha, ~~~~ \psi_{m \alpha} \rightarrow \left( 1 - \frac{A \overline A}{3} \right)^{-\frac{1}{4}} \psi_{m \alpha}  , 
\end{equation}
which must be followed by the gravitino shift
\begin{equation}
\label{shiftoldv} 
\psi_{m \alpha} \rightarrow \psi_{m \alpha} + \frac{i \sqrt{2}}{6} \overline A \left( 1 - \frac{A \overline A}{3}\right)^{-1} \left( \sigma_m \overline\rho \right)_\alpha \, , 
\end{equation}
so that the kinetic mixing terms between the gravitino and the fermion $\rho_\alpha$ are eliminated. 
After integrating out the fermionic auxiliary fields and performing the field redefinitions \eqref{Weylresc} and \eqref{shiftoldv}, the Lagrangian \eqref{Loldv} becomes
\begin{align}
\label{Loldvresc}
\nonumber e^{-1} \mathcal{L} =& -\frac{1}{2} R - \left( 1 - \frac{A \overline A}{3} \right)^{-2} \partial_m A \partial^m \overline A - \frac{i}{2} \left( 1 - \frac{A \overline A}{3} \right)^{-2} \left( \rho \sigma^m \mathcal{D}_m \overline{\rho} + \overline{\rho} {\overline{\sigma}}^m \mathcal{D}_m  \rho \right)
  \\
 \nonumber & + \frac{1}{2} \epsilon^{klmn} \left[ \overline\psi_k \overline\sigma_l \Tilde{\mathcal{D}}_m \psi_n - \psi_k \sigma_l \Tilde{\mathcal{D}}_m \overline\psi_n  + \frac{1}{2} \left( 1 - \frac{A \overline A}{3} \right)^{-1}  \left( A  \partial_k  \overline A - \overline A  \partial_k A  \right) \psi_l \sigma_m \overline\psi_n \right] \\ \nonumber
 & + \frac{i}{12} \left( 1 - \frac{A \overline A}{3} \right)^{-3} \left( A  \partial_m \overline A -  \overline A  \partial_m A \right)  \rho \sigma^m \overline\rho  \\ & - \frac{\sqrt{2}}{2} \left( 1 - \frac{A \overline A}{3} \right)^{-2} \left ( \rho \sigma^m \overline \sigma^n \psi_m \, \partial_n  A + \overline\rho \overline\sigma^m \sigma^n \overline \psi_m \, \partial_n \overline A \right) \\ \nonumber
 & -\mathcal{C} \left( 1 - \frac{A \overline A}{3} \right)^{-\frac{3}{2}}  \bigg[   \left( \psi_m \sigma^{mn} \psi_n + \overline\psi_m \overline\sigma^{mn} \overline\psi_n \right)   + \frac{i\sqrt{2}}{2}  \left( 1 - \frac{A \overline A}{3} \right)^{-1} \left( A \rho \sigma^m \overline\psi_m + \overline A \overline\rho \overline\sigma^m \psi_m \right)  \\
\nonumber & \hspace{0.5cm} + \frac{1}{3} \,  \left( 1 - \frac{A \overline A}{3} \right)^{-2} \left( A^2 \rho \rho + {\overline A}^2 \overline \rho \overline \rho \right) \bigg] + 3 \, \mathcal{C}^2 \left( 1 - \frac{A \overline A}{3} \right)^{-2}  + \text{four-fermion terms} \, .
\end{align}
Thus, the scalar potential is given by
\begin{equation}
\label{Potoldv}
\mathcal{V} \left(A, \overline A \right)  = - 3 \, \mathcal{C}^2 \left( 1 - \frac{A \overline A}{3} \right)^{-2} \,, 
\end{equation}
which has a global maximum at $A=0$, corresponding to an anti-de Sitter vacuum with energy 
\begin{equation}
\label{vevVold}
\langle \mathcal{V} \rangle = \mathcal{V}  \left(0, 0\right) = - 3 \, \mathcal{C}^2 < 0 \, . 
\end{equation}  
Furthermore, by Taylor expanding the potential \eqref{Potoldv} around its maximum point, 
$(A, \overline A)=(0,0)$, we find that the squared mass of the complex scalar $A$ in the standard vacuum \eqref{oldv}
equals
\begin{equation}
\label{mA}
m_A^2 = -2 \, \mathcal{C}^2 < 0 \, , 
\end{equation}
so $A$ is tachyonic. 
From \eqref{vevVold} and \eqref{mA} one can easily see that
\begin{equation}
m_A^2 > \frac{3}{4} \langle \mathcal{V} \,  \rangle \,   , 
\end{equation}
so the tachyon mass satisfies the BF bound and the standard vacuum is a perturbatively stable AdS vacuum.

To deduce whether this vacuum preserves supersymmetry, we must look at the local supersymmetry transformations of the propagating fermionic fields, i.e. the Weyl fermion $\rho_\alpha$ and the gravitino, after the auxiliary fields have been integrated out and the field redefinitions \eqref{Weylresc} and \eqref{shiftoldv} have been performed. These transformations are found to be 
\begin{align}
\label{drhovac}
\delta \rho_\alpha = & \sqrt{2} \, \mathcal{C} \overline A \left( 1 - \frac{A \overline A}{3} \right)^{-\frac{1}{2}} \zeta_\alpha + \dots  \, , \\
\label{dpsivac}
\delta \psi_{m \alpha} = & - 2 \partial_m \zeta_\alpha - \omega_{m a b} (e) \left( \sigma^{ab} \zeta \right)_\alpha -i \, \mathcal{C} \left( 1 - \frac{A \overline A}{3} \right)^{-\frac{3}{2}} \left( \sigma_m \overline\zeta \right)_\alpha + \dots \, ,
\end{align}
where $\omega_{mab}(e)$ denotes the torsion-free spin-connection, the omitted terms involve derivatives of the scalar $A$ or are cubic in the fermions and we have Weyl rescaled the transformation parameters as $\zeta_\alpha \rightarrow \left( 1 - \frac{A \overline A}{3} \right)^{-\frac{1}{4}} \zeta_\alpha$. The fact that $\langle A \rangle = 0$ implies that $\delta \rho_\alpha$ vanishes in the vacuum.  As far as \eqref{dpsivac} is concerned, we first note that in Poincare patch coordinates $(t,x,y,z)$ the AdS$_4$vacuum metric reads
\begin{equation}
\label{AdS}
ds^2 = \frac{1}{\mathcal{C}^2 z^2} \left( -dt^2 + dx^2 + dy^2 + dz^2 \right)   ,  
\end{equation}
which solves the Einstein equation  arising from the Lagrangian \eqref{Loldvresc} if we set $A=\langle A \rangle = 0$ and $\rho_\alpha = \psi_{m \alpha}=0$. Given that the non-vanishing components of the torsion-free spin connection  for  the vielbein ${e_m}^a=\frac{1}{\mathcal{C}z}\delta_m^a$ of AdS$_4$are
\begin{equation}
\label{spinconnection}
{\omega_t}^{03}(e)=-{\omega_t}^{30}(e)={\omega_x}^{13}(e)=-{\omega_x}^{31}(e)={\omega_y}^{23}(e)=-{\omega_y}^{32}(e)=\frac{1}{z} \, ,    
\end{equation}
one can show that $\delta \psi_{m \alpha}$ vanishes in the vacuum if the components of the Weyl spinor transformation parameter $\zeta_\alpha$ are given by
\begin{align}
\label{zeta1}
\zeta_1 =& z^{-\frac{1}{2}} \eta_1 + z^{-\frac{1}{2}} (x - i y) \eta_2 + i (t z^{-\frac{1}{2}} + z^{\frac{1}{2}}) \eta_2^* \, , \\ \label{z2}
\zeta_2 = & i z^{-\frac{1}{2}} \eta_1^* + (t z^{-\frac{1}{2}} - z^{\frac{1}{2}}) \eta_2 + i z^{-\frac{1}{2}} (x + i y) \eta_2^* \,  , \end{align}
where $\eta_1$ and $\eta_2$ are two arbitrary complex constant Grassmann parameters. Therefore, the standard vacuum \eqref{oldv} is supersymmetric. 

Finally, since $\langle A \rangle=0$, the fermion mass terms in this vacuum are just
\begin{equation}
e^{-1} \mathcal{L}_{fermion \, masses} = - \, \mathcal{C} \left( \psi_m \sigma^{mn} \psi_n + \overline\psi_m \overline\sigma^{mn} \overline\psi_n \right)   ,
\end{equation}
so the Weyl fermion $\rho_\alpha$ is massless and the gravitino has Lagrangian mass $m_\psi = \mathcal{C}$ 
and we have the required mass term to describe a 
gravitino with only 2 degrees of freedom in this AdS$_4$ background \cite{Deser:2001us}.

\section{Formulation in terms of chiral superfields}

To further illustrate  the properties of the model \eqref{total-old} 
and relate it to standard literature, 
we will study it now in the dual form. 
First, let us remind the reader of the properties of supersymmetry breaking via nilpotent superfields. 
It is known that standard supergravity coupled to a nilpotent chiral superfield $X$ satisfying 
\be
X^2=0 \,, 
\ee
with K\"ahler potential
\be
K= X \overline X = -3 \ln{\left(1-\frac{X \overline X}{3}\right)} \,, 
\ee
and superpotential
\be
W = f X + {\cal C}\,, 
\ee
where $f$ and $\mathcal{C}$ are real constants, 
breaks supersymmetry. 
To have a model where the nilpotency of $X$ is imposed via a Lagrange multiplier chiral superfield $T$, 
we can consider the following Lagrangian  
\be
\label{Xsugra}
{\cal L}_X =  \int d^2 \Theta \, 2 {\cal E} \ls -3 {\cal R} + T \, X^2 +  f X + {\cal C} - \frac{1}{8} \left( \overline{\cal D}^2 - 8 {\cal R} \right) X \overline X  \rs + c.c. \, 
\ee
Varying \eqref{Xsugra} with respect to $T$ we obtain   
\be
\label{X^2=0}
\delta T \, : \, X^2 = 0 \,,
\ee
while the superspace equation of motion for $X$ reads 
\be
\label{Xeom}
\delta X \, : \,  - \frac{1}{4} \left( \overline{\cal D}^2 - 8 {\cal R} \right) \overline X = -  f - 2 \, T \, X  . 
\ee
Clearly here the nilpotency of $X$ is implemented by the use of a Lagrange multiplier. 
Using these equations of motion we will show that the nilpotent $X$ 
model matches exactly with the full supersymmetry breaking sector of our previous constructions.

We turn once again to the complex linear model \eqref{total-old} and we first recast it in the dual form 
\be
\begin{aligned}
{\cal L}_\text{dual} = & \left[  \int d^2 \Theta \, 2 {\cal E} \left( -3 {\cal R} + {\cal C} \right) + c.c. \right]  
\\
&- \int d^4\theta\, E\ \S \overline{\S}  \ + \int d^4\theta\, E \left( \S \, \Phi + \overline \S \, \overline \Phi \right) 
\\
& + \int d^4\theta \, E \ \frac{1}{64 \, f^2 }\, 
{\cal D}^{\alpha}\S {\cal D}_{\alpha}\S \overline{{\cal D}}_{\dot{\alpha}}\overline{\S}\overline{{\cal D}}^{\dot{\alpha}}\overline{\S}   \,, 
\end{aligned}
\ee
where $\Phi$ is a chiral superfield and $\S$ is now unconstrained. 
Once we integrate out $\Phi$, it will impose \eqref{complin} on $\S$ and we go back to \eqref{total-old}. 
Following \cite{Farakos:2015vba}, we define 
\be
Z = \S - \overline \Phi \, ,
\ee
where $Z$ is an unconstrained superfield, 
and the theory becomes 
\be
\label{Ztheory}
\begin{aligned}
{\cal L}_\text{dual} = & \Big{[} \int d^2 \Theta \, 2 {\cal E} \left( -3 {\cal R} + {\cal C} \right) + c.c. \Big{]}  
+ \int d^4\theta\, E \, \Phi \overline \Phi  
\\
&
- \int d^4\theta\, E\ Z \overline{Z}
+ \int d^4\theta \, E \ \frac{1}{64 \, f^2 }\, 
{\cal D}^{\alpha} Z {\cal D}_{\alpha} Z \overline{{\cal D}}_{\dot{\alpha}}\overline{Z}\overline{{\cal D}}^{\dot{\alpha}}\overline{Z}  . 
\end{aligned}
\ee
Here the supersymmetry breaking sector is only represented by the unconstrained superfield $Z$ and it is decoupled from $\Phi$. 
Therefore, we can study some aspects of the superspace equations of motion of $Z$ independently from $\Phi$. 
Indeed, varying \eqref{Ztheory} with respect to $\overline Z$, 
we obtain the superspace equation of motion  
\be
\label{Zeom}
Z = - \frac{1}{32 f^2} \overline{\cal D}_{\dot{\alpha}} \left( {\cal D}^\alpha Z \, {\cal D}_\alpha Z \, \overline{{\cal D}}^{\dot{\alpha}}\overline{Z}  \right) , 
\ee
which has two solutions that we analyse below. 
These two solutions correspond to two different effective theories that are related to the two different vacua we found for the bosonic sector 
in the previous sections.

The first solution is the trivial 
\be
\label{Z=0}
Z=0
\ee
and corresponds to the supersymmetric vacuum. In this vacuum the Lagrangian \eqref{Ztheory} becomes
\begin{equation}
 {\cal L} =  \Big{[} \int d^2 \Theta \, 2 {\cal E} \left( -3 {\cal R} + {\cal C} \right) + c.c. \Big{]}  + \int d^4\theta\, E \, \Phi \overline \Phi  \, ,    
\end{equation}
so the theory describes a free chiral multiplet $\Phi$
coupled to supergravity with K\"ahler potential  $K= -3 \, \ln \left( 1- \frac{ \Phi \overline \Phi}{3} \right)$ and superpotential $W= \mathcal{C}$. The solution \eqref{Z=0} leads to
\begin{equation}
 \Sigma=\overline\Phi \,,   
\end{equation}
which implies that on-shell
\begin{align}
\lambda_\alpha =  \frac{1}{\sqrt{2}} \mathcal{D}_\alpha \Sigma | &=  \frac{1}{\sqrt{2}} \mathcal{D}_\alpha \overline\Phi | = 0 \, , \\
F =  -\frac{1}{4} \mathcal{D}^2 \Sigma | & = -\frac{1}{4} \mathcal{D}^2 \overline\Phi | = 0 \, , \\ P_{\alpha \dot{\alpha}} =  \overline{\mathcal{D}}_{\dot{\alpha}} \mathcal{D}_\alpha \Sigma| & = \overline{\mathcal{D}}_{\dot{\alpha}} \mathcal{D}_\alpha \overline\Phi | = 0 \, , \\ 
\chi_{\alpha}=\frac{1}{2}\overline {\cal D}_{\dot{\alpha}}{\cal D}_{\alpha}\overline {\cal D}^{\dot{\alpha}}\overline \S| & = \frac{1}{2}\overline {\cal D}_{\dot{\alpha}}{\cal D}_{\alpha}\overline {\cal D}^{\dot{\alpha}} \Phi |=0 \, , 
\end{align}
since $\Phi$ is a chiral superfield.

The second solution, which corresponds to the supersymmetry breaking vacuum, is
\be
\label{Zbreaking}
Z = \frac{1}{2 \sqrt{2} f} \overline{\cal D}^2 \left( X \overline X \right) ,
\ee
where $X$ is a chiral superfield satisfying the on-shell condition \eqref{Xeom} and the nilpotency $X^2=0$. To verify that \eqref{Zbreaking} indeed solves \eqref{Zeom}, we first note that the fact that $X$ is chiral and equations \eqref{X^2=0} and \eqref{Xeom}
allow us to write \eqref{Zbreaking} as 
\begin{equation}
\label{Zbroken}
Z = \sqrt{2} X + \frac{2 \sqrt{2}}{f} \mathcal{R} X \overline X  .
\end{equation}
Furthermore, the nilpotency of $X$ implies
\begin{align}
\label{XDX} X \mathcal{D}_\alpha X  &= 0 \, , \\  
\label{DXDX} \mathcal{D}^\alpha X \mathcal{D}_\alpha X  &= - X \mathcal{D}^2 X  ,
\end{align}
while from \eqref{Xeom} it follows that
\begin{align}
\label{XDDX} X \mathcal{D}^2 X & = 4 f X + 8 \overline T X \overline X , \\
\label{XDDXbar} X \overline{\mathcal{D}}^2 \overline X & = 4 f X + 8 \mathcal{R} X \overline X . 
\end{align}
By making repeated use of equations \eqref{X^2=0}, \eqref{XDX}, \eqref{DXDX} and \eqref{XDDX}, we can show that \eqref{Zbroken} satisfies 
\begin{align*}
\mathcal{D}^\alpha Z \mathcal{D}_\alpha Z & = \left( 2 + \frac{8}{f} \mathcal{R} \overline X \right) \mathcal{D}^\alpha X \mathcal{D}_\alpha X = - \left( 2 + \frac{8}{f} \mathcal{R} \overline X \right)  X \mathcal{D}^2 X \\ 
&= - \left( 2 + \frac{8}{f} \mathcal{R} \overline X \right)   \left( 4 f X + 8 \overline T X \overline X \right)  = - 8 \left ( f + 2 \overline T \overline X + 4 \mathcal{R} \overline X \right) X \, , 
\end{align*}
thus 
\begin{align*}
  \mathcal{D}^\alpha Z \mathcal{D}_\alpha Z \overline{\mathcal{D}}^{\dot{\alpha}} \overline Z &=   - 8 \left ( f + 2 \overline T \overline X + 4 \mathcal{R} \overline X \right) X  \left[ \sqrt{2} \overline{\mathcal{D}}^{\dot{\alpha}}  \overline X + \frac{2 \sqrt{2}}{f} \left( \overline{\mathcal{D}}^{\dot{\alpha}} \overline{\mathcal{R}}\right) \overline X X + \frac{2 \sqrt{2}}{f} \overline{\mathcal{R}}  \left( \overline{\mathcal{D}}^{\dot{\alpha}} \overline X \right) X  \right] \\
  & \overset{\eqref{X^2=0}}{=} - 8 \sqrt{2} \left( f + 2 \overline T \overline X + 4 \mathcal{R} \overline X \right) X \overline{\mathcal{D}}^{\dot{\alpha}}  \overline X  \overset{\eqref{XDX}}{=} - 8 \sqrt{2} f X \overline{\mathcal{D}}^{\dot{\alpha}}  \overline X \, , 
\end{align*}
so if $Z$ is given by \eqref{Zbreaking}, the right hand side of \eqref{Zeom} equals
\begin{align*}
- \frac{1}{32 f^2} \overline{\cal D}_{\dot{\alpha}} \left( {\cal D}^\alpha Z \, {\cal D}_\alpha Z \, \overline{{\cal D}}^{\dot{\alpha}}\overline{Z}  \right) & = \frac{\sqrt{2}}{4f} \overline{\cal D}_{\dot{\alpha}} \left( X \overline{\cal D}^{\dot{\alpha}} \overline X \right)   =  \frac{\sqrt{2}}{4f} X \overline{\mathcal{D}}^2 \overline X \\
&\overset{\eqref{XDDXbar}} {=}  \sqrt{2} X + \frac{2 \sqrt{2}}{f} \mathcal{R} X \overline X = Z \, , 
\end{align*}
which concludes the proof that \eqref{Zbreaking}
is a solution of the equation of motion \eqref{Zeom}.

We conclude that in the supersymmetry breaking vacuum the $Z$ sector describes only a propagating goldstino 
and an effective description of the $Z$ sector in \eqref{Ztheory} is given precisely by the $X$ sector of \eqref{Xsugra} 
because both Lagrangian descriptions give rise to equations of motion which describe exactly the same propagating degrees of freedom coupled to supergravity in the same way. 
In particular, 
in the new vacuum \eqref{newv} an on-shell description of the model \eqref{total-old} 
is provided by standard supergravity coupled to a nilpotent chiral superfield 
\be
\label{nilX}
X = \frac{G^2}{2 F^X} + \sqrt 2 \Theta G + \Theta^2 F^X \,, 
\ee
and a free chiral superfield  
\be
\label{Phi}
\Phi = A^{\Phi} + \sqrt 2 \Theta \psi + \Theta^2 F^\Phi \, . 
\ee
The K\"ahler potential is 
\be
\label{KK}
K= -3 \, \ln \left( 1- \frac{ \Phi \overline \Phi}{3} - \frac{X \overline X}{3} \right) \,, 
\ee
and the superpotential is 
\be
\label{WW}
W = f X + {\cal C} . 
\ee 
We can then deduce the on-shell component fields of the $\Sigma$ multiplet from the relation 
\begin{equation}
\label{Sigma=Z+Phi}
\Sigma = Z + \overline\Phi =   \sqrt{2} X + \frac{2 \sqrt{2}}{f} \mathcal{R} X \overline X + \overline\Phi  \, , 
\end{equation} 
which verifies that the full component field solution we presented in the third section exists and is self-consistent. 
We have that 
\begin{align}
\label{A} A  = \Sigma |=  &  {\overline{A}}^\Phi + \frac{G^2}{\sqrt{2} F^X} - \frac{\sqrt{2}}{12 f} M \frac{G^2 {\overline G}^2}{F^X {\overline F}^X} \, , \\
\nonumber \lambda_\alpha = \frac{1}{\sqrt{2}} \mathcal{D}_\alpha \Sigma | = & \mathcal{D}_\alpha X | + \frac{2}{f} \big{[} \left( \mathcal{D}_\alpha \mathcal{R} \right) X + \mathcal{R} \left(\mathcal{D}_\alpha X\right) \big{]} \overline X | \\ 
= &\label{lambda} \sqrt{2} G_\alpha - \frac{\sqrt{2}}{6f} M \frac{ {\overline G}^2}{ {\overline F}^X} G_\alpha + \frac{1}{2 f} \frac{G^2 {\overline G}^2}{F^X {\overline F}^X} \mathcal{D}_\alpha \mathcal{R} | \, , \\
 \label{rho} \rho_\alpha = \frac{1}{\sqrt{2}} \mathcal{D}_\alpha \overline\Sigma |  = & \frac{1}{\sqrt{2}} \mathcal{D}_\alpha \Phi | + \frac{2}{f} \overline{\mathcal{R} } \overline X \mathcal{D}_\alpha X | = \psi_\alpha - \frac{\sqrt{2}}{6 f} \overline M \frac{ {\overline G}^2}{ {\overline F}^X} G_\alpha \, , \\
 \nonumber F = - \frac{1}{4} \mathcal{D}^2 \Sigma |  = & -\frac{\sqrt{2}}{4} \mathcal{D}^2 X | -\frac{\sqrt{2}}{2f} \left( X \overline X \mathcal{D}^2 \mathcal{R} + 2 \overline X \mathcal{D}^\alpha X \mathcal{D}_\alpha \mathcal{R} +  \mathcal{R} \overline X \mathcal{D}^2 X \right) | \\ \label{F}
 = &  \sqrt{2} F^X - \frac{\sqrt{2}}{6 f} M \frac{ {\overline G}^2}{ {\overline F}^X} F^X - \frac{1}{f} \frac{ {\overline G}^2}{ {\overline F}^X} G^\alpha \mathcal{D}_\alpha \mathcal{R} | - \frac{\sqrt{2}}{8 f} \frac{G^2 {\overline G}^2}{F^X {\overline F}^X} \mathcal{D}^2 \mathcal{R}| \, , \\ \nonumber
 P_{\alpha \dot{\alpha}}  = \overline{\mathcal{D}}_{\dot{\alpha}} \mathcal{D}_\alpha \Sigma |   = & \sqrt{2} \overline{\mathcal{D}}_{\dot{\alpha}} \mathcal{D}_\alpha X | + \frac{2 \sqrt{2}}{f} \big{[} \left( \overline{\mathcal{D}}_{\dot{\alpha}} \mathcal{D}_\alpha \mathcal{R}\right) X \overline X - \left( \mathcal{D}_\alpha \mathcal{R}\right) X \overline{\mathcal{D}}_{\dot{\alpha}} \overline X  \\ \nonumber
 & + \mathcal{R} \left( \overline{\mathcal{D}}_{\dot{\alpha}} \mathcal{D}_\alpha X\right) \overline X - \mathcal{R} \mathcal{D}_\alpha X \overline{\mathcal{D}}_{\dot{\alpha}} \overline X\big{]} | \\ \nonumber
  = &- 2 i \sqrt{2} \left( \sigma^m \right)_{\alpha \dot{\alpha}} \hat{D}_m \left( \frac{G^2}{2 F^X} \right) + \frac{2 \sqrt{2}}{3 f} M G_\alpha {\overline G}_{\dot{\alpha}} + \frac{2}{f} \frac{G^2}{F^X} {\overline G}_{\dot{\alpha}} \mathcal{D}_\alpha \mathcal{R} | \\ \label{P}
& + \frac{i \sqrt{2}}{3 f} M \frac{{\overline G}^2}{{\overline F}^X} \left( \sigma^m \right)_{\alpha \dot{\alpha}} \hat{D}_m \left( \frac{G^2}{2 F^X} \right) + \frac{\sqrt{2}}{2 f}   \frac{G^2 {\overline G}^2}{F^X {\overline F}^X} \overline{\mathcal{D}}_{\dot{\alpha}} \mathcal{D}_\alpha \mathcal{R}| \, ,  \\
\nonumber 
\chi_{\alpha}=\frac{1}{2}\overline {\cal D}_{\dot{\alpha}}{\cal D}_{\alpha}\overline {\cal D}^{\dot{\alpha}}\overline \S| = & \frac{1}{\sqrt{2}} \overline {\cal D}_{\dot{\alpha}}{\cal D}_{\alpha}\overline {\cal D}^{\dot{\alpha}} \overline X | + \frac{\sqrt{2}}{f} \big{[} \left(  \overline {\cal D}_{\dot{\alpha}}{\cal D}_{\alpha}\overline {\cal D}^{\dot{\alpha}} \overline{\mathcal{R}} \right) \overline X X + \left({\cal D}_{\alpha}\overline {\cal D}^{\dot{\alpha}} \overline{\mathcal{R}}\right) \left(\overline {\cal D}_{\dot{\alpha}} \overline X  \right) X \\
\nonumber & - \left(\overline{\mathcal{D}}^2 \overline{\mathcal{R}}\right) \overline X \mathcal{D}_\alpha X -2 \left(\overline{\cal D}_{\dot{\alpha}} \overline{\mathcal{R}} \right) \left( \overline {\cal D}^{\dot{\alpha}} \overline X \right) \mathcal{D}_\alpha X + \left( \overline{\cal D}^{\dot{\alpha}} \overline{\mathcal{R}} \right) \overline X \overline{\cal D}_{\dot{\alpha}} \mathcal{D}_\alpha X \\
\nonumber & + \left( \overline{\cal D}_{\dot{\alpha}} \overline{\mathcal{R}} \right) \left( \mathcal{D}_\alpha \overline{\cal D}^{\dot{\alpha}} \overline X\right) X + \overline{\mathcal{R}} \left(\overline {\cal D}_{\dot{\alpha}}{\cal D}_{\alpha}\overline {\cal D}^{\dot{\alpha}} \overline X \right) X - \overline{\mathcal{R}} \left( \overline {\cal D}^2 \overline X
\right) \mathcal{D}_\alpha X \\
\nonumber & + \overline{\mathcal{R}} \left( \overline {\cal D}^{\dot{\alpha}} \overline X \right) \overline {\cal D}_{\dot{\alpha}} \mathcal{D}_\alpha X \big{]} | \\
\nonumber =& \,  2 i \left( \sigma^m \right)_{\alpha \dot{\alpha}} \hat{D}_m {\overline G}^{\dot{\alpha}} + \frac{5}{3} b_{\alpha \dot{\alpha}} {\overline G}^{\dot{\alpha}} - \frac{4}{3 f} \overline M {\overline F}^X G_\alpha \\
\nonumber & - \frac{i}{3 f} \overline M \frac{G^2}{F^X} \left( \sigma^m \right)_{\alpha \dot{\alpha}} \hat{D}_m {\overline G}^{\dot{\alpha}}  + \frac{2i}{3f} \overline M \hat{D}_m \left( \frac{G^2}{2 F^X} \right) \left( \sigma^m \overline  G \right)_\alpha - \frac{5}{18 f} \overline M \frac{G^2}{F^X} b_{\alpha \dot{\alpha}} {\overline G}^{\dot{\alpha}} \\ \nonumber & - \frac{4 \sqrt{2}}{f} G_\alpha {\overline G}_{\dot{\alpha}} \overline {\cal D}^{\dot{\alpha}} \overline{\mathcal{R}} |  - \frac{1}{f} \frac{G^2}{F^X} {\overline G}^{\dot{\alpha}} \mathcal{D}_\alpha \overline {\cal D}_{\dot{\alpha}} \overline{\mathcal{R}}| - \frac{1}{f} \frac{{\overline G}^2}{{\overline F}^X} G_\alpha \overline{\mathcal{D}}^2 \overline{\mathcal{R}}| \\ \nonumber
& - \frac{i \sqrt{2}}{f} \frac{{\overline G}^2}{{\overline F}^X} \left( \sigma^m \right)_{\alpha \dot{\alpha}} \hat{D}_m \left( \frac{G^2}{2 F^X} \right) \overline {\cal D}^{\dot{\alpha}} \overline{\mathcal{R}}| + \frac{i \sqrt{2}}{f} \frac{G^2}{F^X} \left( \sigma^m \right)_{\alpha \dot{\alpha}} \hat{D}_m \left( \frac{{\overline G}^2}{2 {\overline F}^X} \right) \overline {\cal D}^{\dot{\alpha}} \overline{\mathcal{R}}| \\
& + \frac{\sqrt{2}}{4f} \frac{G^2 {\overline G}^2}{F^X {\overline F}^X} \overline {\cal D}_{\dot{\alpha}}{\cal D}_{\alpha}\overline {\cal D}^{\dot{\alpha}} \overline{\mathcal{R}} | \, , 
\end{align}
where
\begin{align}
\hat{D}_m \left( \frac{G^2}{2 F^X} \right) \equiv & \, \partial_m \left( \frac{G^2}{2 F^X} \right) - \frac{1}{\sqrt{2}} \psi_m G   \, ,
\\
\hat{D}_m G_\alpha \equiv & \, \mathcal{D}_m G_\alpha - \frac{1}{\sqrt{2}} \psi_{m \alpha} F^X - \frac{i}{\sqrt{2}} \left( \sigma^n \overline{\psi}_m 
\right)_\alpha \hat{D}_n \left( \frac{G^2}{2 F^X} \right)  , \\
\overline{\mathcal{D}}_{\dot{\alpha}} \mathcal{D}_\alpha \mathcal{R} | = & \, i \left( \sigma^m \right)_{\alpha \dot{\alpha}} \left( \frac{1}{3} \partial_m M + {\psi_m}^\beta \mathcal{D}_\beta \mathcal{R} |\right) , \\
\overline {\cal D}_{\dot{\alpha}}{\cal D}_{\alpha}\overline {\cal D}^{\dot{\alpha}} \overline{\mathcal{R}} | = & \, 2i \left( \sigma^m \right)_{\alpha \dot{\alpha}} \left[ \mathcal{D}_m\left( \overline{\mathcal{D}}^{\dot{\alpha}} \overline{\mathcal{R}} | \right) - \frac{1}{2} {\psi_m}^\beta \mathcal{D}_\beta \overline{\mathcal{D}}^{\dot{\alpha}} \overline{\mathcal{R}}| + \frac{1}{4} {\overline\psi_m}^{\dot{\alpha}} \overline{\mathcal{D}}^2 \overline{\mathcal{R}}| \right] \\ \nonumber
& + \frac{5}{3} b_{\alpha \dot{\alpha}} \overline{\mathcal{D}}^{\dot{\alpha}} \overline{\mathcal{R}}| , 
\end{align}
and we refer the reader to \cite{Wess:1992cp} for the expressions for $\mathcal{D}_\alpha \mathcal{R}|$ and $\mathcal{D}^2 \mathcal{R}|$.

Furthermore, using standard supergravity formulae that can be found in \cite{Wess:1992cp} one can write down the component form of the Lagrangian for the theory of the  chiral multiplets \eqref{nilX} and \eqref{Phi}
coupled to supergravity with K\"ahler potential and superpotential given by \eqref{KK} and \eqref{WW} respectively. The superspace form of this Lagrangian reads
\begin{equation}
\label{LXPhi}
\mathcal{L}_{X,\Phi} = \int d^4\theta\, E \left( \Phi \overline \Phi + X \overline X - 3 \right)+ \Big{[} \int d^2 \Theta \, 2 {\cal E} \left( f X + {\cal C} \right) + c.c. \Big{]}   \, 
\end{equation}
 After integrating out the auxiliary fields $M$, $b_m$, $F^\Phi$ and $F^X$, \eqref{LXPhi} becomes
\begin{align}
\nonumber e^{-1} \mathcal{L}_{X,\Phi} = & \frac{1}{6} \left( A^\Phi {\overline A}^\Phi - 3  \right) R - \partial_m A^\Phi \, \partial^m {\overline A}^\Phi  - f^2 + 3 \, \mathcal{C}^2 \\
\nonumber & -\frac{i}{2} \left(\psi \sigma^m \mathcal{D}_m \overline{\psi} + G \sigma^m \mathcal{D}_m \overline{G}+ \overline{\psi} {\overline{\sigma}}^m \mathcal{D}_m \psi  + \overline{G} {\overline{\sigma}}^m \mathcal{D}_m G \right) \\
\nonumber & - \frac{1}{12} \left( A^\Phi  {\overline A}^\Phi -3 \right) \epsilon^{klmn} \left( \overline\psi_k \overline\sigma_l \psi_{mn} - \psi_k \sigma_l \overline\psi_{mn}  \right)
+ \frac{\sqrt{2}}{3} \left( {\overline A}^\Phi  \psi \sigma^{mn} \psi_{mn} +  A^\Phi \overline{\psi} {\overline{\sigma}}^{mn} \overline{\psi}_{mn}\right) \\ \nonumber
&-\frac{\sqrt{2}}{2} \left( \psi_n \sigma^m \overline{\sigma}^n \psi \, \partial_m {\overline A}^\Phi + {\overline{\psi}}_n \overline{\sigma}^m \sigma^n \overline{\psi} \, \partial_m A^\Phi \right)  + \frac{1}{4} \epsilon^{klmn} \left( {\overline A}^\Phi \partial_k A^\Phi - A^\Phi  \partial_k {\overline A}^\Phi \right) \psi_l \sigma_m {\overline{\psi}}_n \\ \nonumber
& - \frac{i}{\sqrt{2}} f \left( G \sigma^m \overline\psi_m - \psi_m \sigma^m \overline G \right) -  \mathcal{C} \left( GG + \overline G \overline G \right) - \mathcal{C} \left ( \psi_m \sigma^{mn} \psi_n + \overline\psi_m \overline\sigma^{mn} \overline\psi_n \right) \\ \label{Ldualonshell} 
& - \frac{1}{4} \left( A^\Phi {\overline A}^\Phi  - 3 \right)^{-1} \left( {\overline A}^\Phi \partial_m A^\Phi - A^\Phi  \partial_m {\overline A}^\Phi  \right) \Big{[} {\overline A}^\Phi \partial^m A^\Phi - A^\Phi \partial^m {\overline A}^\Phi  \\ \nonumber 
& \hspace{0.35cm} - i \psi \sigma^m \overline\psi -i G \sigma^m \overline G 
 - \sqrt{2} \left( {\overline A}^\Phi \psi^m \psi -A^\Phi  \overline\psi^m \overline\psi \right) \Big{]} \\ \nonumber 
 &+ \text{terms at least quartic in the fermions} \, .   
\end{align}
Using equations \eqref{A}-\eqref{rho} and the fact that on-shell, $F^X = -f + \text{fermions}$, one can show that the Lagrangians \eqref{Lonshell}, which describes the theory \eqref{total-old} around the supersymmetry breaking vacuum, and \eqref{Ldualonshell} coincide at quadratic order in the fermions, which once more verifies our previous results. Of course, the duality procedure that was carried out implies that these two Lagrangians should be exactly equal given equations \eqref{A}-\eqref{rho} with the auxiliary fields involved replaced by the solutions of the corresponding equations of motion, since the theory \eqref{total-old} around the new vacuum \eqref{newv} and the one described by \eqref{LXPhi} have been shown to be on-shell equivalent.

In conclusion, our analysis shows that the theory \eqref{total-old} describes two backgrounds: 
\vspace*{-5mm}
\begin{enumerate}
\setlength\itemsep{0.2em}
\item A supersymmetric background  with a free chiral superfield $\Phi$ coupled to supergravity with K\"ahler potential $K= -3 \, \ln \left( 1- \frac{ \Phi \overline \Phi}{3} \right)$  and superpotential $W={\cal C}$. 
\item A non-supersymmetric one with a nilpotent chiral superfield $X$ and a chiral superfield $\Phi$ coupled to supergravity with K\"ahler potential \eqref{KK} and superpotential \eqref{WW}. 

\end{enumerate}

\section{Non-standard kinetic function and higher order terms} 

In this section we want to study the properties of models that have a non-standard kinetic function which 
violates the condition \eqref{Pcondition} and are also coupled to the higher order terms. 
Since this analysis has not been explicitly performed in global supersymmetry earlier, 
we will first discuss the global SUSY case and then turn to supergravity.

We therefore consider the following globally supersymmetric theory
\begin{equation}
\label{global1}
\mathcal{L} = - \int d^4 \theta \, g(\Sigma + \overline\Sigma)   + \frac{1}{64 f^2}  \int d^4 \theta \, D^\alpha \Sigma D_\alpha \Sigma {\overline D}_{\dot{\alpha}} \overline\Sigma {\overline D}^{\dot{\alpha}} \overline\Sigma \, ,
\end{equation}
where $\Sigma$ is a complex linear superfield,
\begin{equation}
{\overline D}^2  \Sigma = 0 
\end{equation}
and $g$ is an arbitrary real function of a single variable. Clearly, in \eqref{global1} the superspace kinetic function is 
\begin{equation}
\Omega(\Sigma, \overline{\Sigma}) = g(\Sigma+\overline{\Sigma}) , 
\end{equation}
which does not satisfy \eqref{Pcondition}, as we pointed out in section 2. In terms of the component fields of $\Sigma$, which are defined in \cite{Farakos:2013zsa}, the bosonic sector of \eqref{global1} reads
\begin{align}
 \label{bosglobal1}
\mathcal{L}^{(b)}=&-g'' (\partial_m A \partial^m A + \partial_m A  \partial^m \overline A + \partial_m \overline A  \partial^m \overline A) 
 + ig'' (P^m \partial_m A - {\overline P}^m \partial_m \overline A) \nonumber \\
 & + \frac{1}{4} g'' \left( P_m P^m + 2 P_m {\overline P}^m + {\overline P}_m  {\overline P}^m   \right) - g'' F \overline F \\  \nonumber
 & + \frac{1}{64 f^2}\left( P_m P^m \overline{P}_n \overline{P}^n 
-8P_m\overline{P}^mF\overline{F}+16   F^2\overline{F}^2\right),
\end{align}
where we have introduced the notation 
\begin{align}
g & = g(A+\overline{A}) = g(\Sigma+\overline{\Sigma})| , \\
g' & = g'(A+\overline{A}) = \frac{dg(u)}{du}\bigg{|}_{u=A+\overline{A}} , \\
g'' & = g''(A+\overline{A}) = \frac{d^2g(u)}{du^2}\bigg{|}_{u=A+\overline{A}} \,. 
\end{align}
The equation of motion for $F$ reads
\begin{equation}
\label{Fglobal1}
F \left( g'' + \frac{1}{8 f^2} P_m {\overline P}^m - \frac{1}{2 f^2} F \overline F \right) = 0  
\end{equation}
and has clearly two solutions 
\begin{align}
\label{Fglobal=01}
F = & \,  0 \, ,\\
\label{nonzeroF1}
F \overline F = &\,  2 f^2 g'' + \frac{1}{4} P_m {\overline P}^m.
\end{align}
Substituting the trivial solution \eqref{Fglobal=01} into \eqref{bosglobal1} we obtain 
\be
\begin{aligned}
\label{LF=01}
\mathcal{L}^{(b)}_{F=0}=&-g''(\partial_m A \partial^m A + \partial_m A  \partial^m \overline A + \partial_m \overline A  \partial^m \overline A )
 + ig'' (P^m \partial_m A - {\overline P}^m \partial_m \overline A)  \\
 & + \frac{1}{4} g'' \left( P_m P^m + 2 P_m {\overline P}^m + {\overline P}_m  {\overline P}^m   \right)  
  + \frac{1}{64 f^2} P_m P^m \overline{P}_n \overline{P}^n , 
\end{aligned}
\ee
so the equation of motion for $P_m$ reads
\begin{equation}
\label{Pglobal1}
g''(P_m + {\overline P}_m) = - 2 i g'' \partial_m A - \frac{1}{16 f^2} P_m {\overline P}_n  {\overline P}^n \,, 
\end{equation} 
which implies that 
\begin{equation}
\label{d_mA1}
\partial_m \overline A = - \partial_m A + \frac{i}{32 f^2 g''} P_m {\overline P}_n {\overline P}^n - \frac{i}{32 f^2 g''} {\overline P}_m P_n P^n \,. 
\end{equation}
Plugging \eqref{d_mA1} into \eqref{LF=01} and using \eqref{Pglobal1} we get 
\begin{equation}
   \mathcal{L}^{(b)}_{F=0} = 0 + \text{terms with four or more vectors}. 
\end{equation}
On the other hand, the solution \eqref{nonzeroF1} leads to the bosonic sector 
\begin{align}
\label{LnonzeroF1} 
\mathcal{L}^{(b)}_{F \overline F\ne 0}=& -g'' (\partial_m A \partial^m A + \partial_m A  \partial^m \overline A + \partial_m \overline A  \partial^m \overline A )
 + i g''(P^m \partial_m A - {\overline P}^m \partial_m \overline A) \\
 & + \frac{1}{4}g'' \left( P_m P^m +  P_m {\overline P}^m + {\overline P}_m  {\overline P}^m   \right) -  f^2 (g'')^2 + \frac{1}{64f^2} \left( P_m P^m {\overline P}_n {\overline P}^n - P_m {\overline P}^m  P_n {\overline P}^n \right) ,\nonumber
 \end{align}
so the equation of motion for $P_m$ is
\begin{equation}
\label{PnonzeroF1}
g''\left(P_m + \frac{1}{2} {\overline P}_m\right) = -2i g'' \partial_m A - \frac{1}{16 f^2} \left( {\overline P}_n {\overline P}^n P_m - P_n {\overline P}^n {\overline P}_m \right),
\end{equation}
which implies that 
\begin{equation}
\label{Psol1}
P_m = - \frac{4i}{3} \left( 2 \partial_m A + \partial_m \overline A \right) + \text{terms cubic in $P$}.
\end{equation}
Substituting \eqref{Psol1} into \eqref{LnonzeroF1} gives
\be
\begin{aligned}
    \mathcal{L}^{(b)}_{F \overline F\ne 0}= &\frac{1}{3} g'' \left( \partial_m A \partial^m A + \partial_m A  \partial^m \overline A + \partial_m \overline A  \partial^m \overline A\right) - f^2 (g'')^2 \\
    & + \text{terms with four or more vectors}\,.
\end{aligned}    
\ee
If we write $A=A_1+iA_2$, where $A_1$ and $A_2$ are real scalar fields, the scalar kinetic terms read 
\begin{equation}
\mathcal{L}_{\text{kin, scalars}} = g'' \partial_m A_1 \partial^m A_1 - \frac{1}{3}g'' \partial_m A_2 \partial^m A_2 \, .
\end{equation}
We see that the kinetic terms for $A_1$ and $A_2$ have always opposite signs, thus exactly one of them is a ghost.

We now consider the coupling of the model \eqref{global1} to supergravity, which is described by the superspace Lagrangian 
\begin{equation}
\label{Llocal1}
 \mathcal{L} = - \int d^4 \theta E \, g(\Sigma + \overline\Sigma) -3\left(  \int d^2 \Theta \, 2 {\cal E}   {\cal R} + c.c. \right)   + \mathcal{L}_{\text{HD}} \, ,    
\end{equation}
where $\Sigma$ is a complex linear superfield, $ \mathcal{L}_{\text{HD}}$ is given by \eqref{ep1} 
and we have now set the gravitino mass parameter $\mathcal{C}$ equal to zero such that our analysis remains tractable. 
The bosonic sector of \eqref{Llocal1} is given by 
\be
\label{Lblocal1}
\begin{aligned}
 e^{-1} \mathcal{L}^{(b)} =& \frac{1}{6} \left[ \left(A + \overline A\right)g' - g - 3 \right] R - g'' (\partial_m A \partial^m A + \partial_m A \partial^m \overline A + \partial_m \overline A \partial^m \overline A ) \\
& + ig'' (P^m \partial_m A - \overline{P}^m \partial_m \overline A) + \frac{1}{4} g''\left( P_m P^m + 2 P_m {\overline P}^m + {\overline P}_m {\overline P}^m \right) \\
& + \frac{i}{3} g'' b^m \left( \overline A  - A   \right)\partial_m\left( A + \overline A \right) - \frac{1}{9} \left[\left(A + \overline A\right)g' - g - 3 \right] b_m b^m\\
& - g'' F \overline F - \frac{1}{3} g'' \left( A M F + \overline A \overline M \overline F \right) + \frac{1}{9}\left[ \left(A + \overline A\right)g' - A \overline{A}g''- g - 3 \right]M \overline M \\
& + \frac{1}{64  f^2}\left( P_m P^m \overline{P}_n \overline{P}^n 
-8P_m\overline{P}^mF\overline{F}+16   F^2\overline{F}^2\right) \,. 
\end{aligned}
\ee
After integrating out the auxiliary fields $M$ and $b_m$, \eqref{Lblocal1} becomes 
\be
\label{LbMbintout1}
\begin{aligned}
e^{-1} \mathcal{L}^{(b)}   =& \frac{1}{6} \left[\left(A + \overline A\right)g' - g - 3 \right] R - g'' (\partial_m A \partial^m A + \partial_m A \partial^m \overline A + \partial_m \overline A \partial^m \overline A ) \\
& + i g''(P^m \partial_m A - \overline{P}^m \partial_m \overline A) + \frac{1}{4} g'' \left( P_m P^m + 2 P_m {\overline P}^m + {\overline P}_m {\overline P}^m \right) \\
& - \frac{1}{4} (g'')^2\left[ \left(A + \overline A\right)g'-g- 3 \right]^{-1} (\overline A - A )^2 \partial_m (A + \overline A) \partial^m (A + \overline A) \\
& - g'' F \overline F - (g'')^2 A \overline{A}\left[ \left(A + \overline A\right)g'-A \overline{A} g''-g- 3 \right]^{-1}  F \overline F  \\
& + \frac{1}{64 f^2}\left( P_m P^m {\overline P}_n {\overline P}^n  - 8 P_m {\overline P}^m F \overline F + 16 F^2 {\overline F}^2 \right), 
 \end{aligned}
\ee
so the equation of motion for the auxiliary field $F$ reads 
\begin{equation}
\label{Feqom1}
F \left\{ g''+ (g'')^2 A \overline{A}\left[ \left(A + \overline A\right)g'-A \overline{A} g''-g- 3 \right]^{-1} + \frac{1}{8 f^2}  P_m {\overline P}^m - \frac{1}{2 f^2} F \overline F  \right\} = 0 
\end{equation}
and has two solutions 
\begin{align}
\label{trivialF1}
F=&\,0 \, , \\
\label{nontrivialF1}
F \overline F=&\, 2f^2 g'' + 2 f^2 (g'')^2 A \overline{A} \left[ \left(A + \overline A\right)g'-A \overline{A} g''-g- 3 \right]^{-1} + \frac{1}{4} P_m {\overline{P}}^m \,. 
\end{align}
The trivial solution \eqref{trivialF1} gives
\be
\begin{aligned}
\label{LbMF=0intout1}
e^{-1} \mathcal{L}^{(b)}   =& \frac{1}{6} \left[ \left(A + \overline A\right)g'-g - 3 \right] R - g'' (\partial_m A \partial^m A + \partial_m A \partial^m \overline A + \partial_m \overline A \partial^m \overline A ) \\
& + i g'' (P^m \partial_m A - \overline{P}^m \partial_m \overline A) + \frac{1}{4} g'' \left( P_m P^m + 2 P_m {\overline P}^m + {\overline P}_m {\overline P}^m \right) \\
& -\frac{1}{4} (g'')^2\left[ \left(A + \overline A\right)g'-g- 3 \right]^{-1} (\overline A - A )^2 \partial_m (A + \overline A) \partial^m (A + \overline A)  \\
& + \frac{1}{64 f^2} P_m P^m {\overline P}_n {\overline P}^n  .
\end{aligned}
\ee
Then, the equation of motion for $P_m$ is
\begin{equation}
\label{Peqom1}
g''(P_m +{\overline P}_m) = - 2i g'' \partial_m A - \frac{1}{16 f^2} P_m {\overline P}_n {\overline P}^n 
\end{equation}
and implies that
\begin{equation}
\label{dAlocal1}
\partial_m \overline A = - \partial_m A  + \frac{i}{32 f^2 g''} P_m {\overline P}_n {\overline P}^n - \frac{i}{32 f^2 g''} {\overline P}_m P_n P^n \,. 
\end{equation}
Substituting \eqref{dAlocal1} into \eqref{LbMF=0intout1} and using \eqref{Peqom1} we get
\begin{equation}
 e^{-1} \mathcal{L}^{(b)}   =    \frac{1}{6} \left[ \left(A + \overline{A} \right)g' - g -3 \right] R + \text{terms with four or more vectors}
\end{equation}
After performing the Weyl rescaling
\begin{equation}
\label{WeylReA1}
    {e_m}^a \rightarrow  \left[ 1 + \frac{1}{3}g -\frac{1}{3}\left( A+\overline{A}\right)g' \right]^{-\frac{1}{2}}   {e_m}^a
\end{equation}
and using \eqref{dAlocal1} we are left with
\begin{equation}
     e^{-1} \mathcal{L}^{(b)} = - \frac{1}{2} R + \text{terms with four or more vectors}. 
 \end{equation}
On the other hand, the solution \eqref{nontrivialF1} of the equation of motion for $F$ leads to the bosonic sector \be
\begin{aligned}
\label{LbMbFnonzero1}
e^{-1} \mathcal{L}^{(b)} =& \frac{1}{6} \left[ \left(A + \overline A\right)g'-g - 3 \right] R - g'' (\partial_m A \partial^m A + \partial_m A \partial^m \overline A + \partial_m \overline A \partial^m \overline A ) \\
& -\frac{1}{4} (g'')^2\left[ \left(A + \overline A\right)g'-g- 3 \right]^{-1} (\overline A - A )^2 \partial_m (A + \overline A) \partial^m (A + \overline A)  \\
& + i g'' (P^m \partial_m A - \overline{P}^m \partial_m \overline A) + \frac{1}{4} g'' \left( P_m P^m +  P_m {\overline P}^m + {\overline P}_m {\overline P}^m \right) \\
&  -\frac{1}{4} (g'')^2 A \overline{A} \left[\left(A + \overline A\right)g'-A \overline{A} g''-g - 3 \right]^{-1} P_m {\overline{P}}^m\\
& - f^2 (g'')^2 \{ 1 + g''A \overline{A} [(A+\overline{A})g'-A \overline{A} g''-g - 3]^{-1}\}^2                   \\
& + \frac{1}{64 f^2} \left( P_m P^m {\overline P}_n {\overline P}^n  - P_m {\overline{P}}^m P_n {\overline{P}}^n \right), 
\end{aligned}
\ee
so the equation of motion for $P_m$ reads
\begin{align}
\label{P'eqom1}
& g'' P_m + \frac{1}{2} g'' \left\{ 1 - g''A \overline{A} \left[\left(A+\overline{A}\right)g'-A \overline{A} g''-g - 3\right]^{-1}\right\} {\overline P}_m = \nonumber
\\& - 2i g'' \partial_m A - \frac{1}{16 f^2} \left({\overline P}_n {\overline P}^n P_m  -   P_n {\overline P}^n {\overline P}_m\right), 
\end{align}
which implies 
\be
\label{Psol2}
\begin{aligned}
P_m = & -4i \{ 3 + 2g''A \overline{A} [(A+\overline{A})g'-A \overline{A} g''-g - 3]^{-1} \\ 
&\hspace{0.8cm}- (g'')^2(A \overline{A})^2 [(A+\overline{A})g'-A \overline{A} g''-g - 3]^{-2}\}^{-1} \times
\\ & \{ 2 \partial_m A + \{1-g''A \overline{A}[(A+\overline{A})g'-A \overline{A} g''-g - 3]^{-1} \} \partial_m \overline{A} \} \\
& + \text{terms cubic in $P$}.
\end{aligned}
\ee
After substituting \eqref{Psol2} into \eqref{LbMbFnonzero1} and performing the Weyl rescaling \eqref{WeylReA1}, \eqref{LbMbFnonzero1} becomes 
\be 
\begin{aligned}
\label{Llocalonshell}
e^{-1} \mathcal{L}^{(b)} =& - \frac{1}{2} R - p(A, \overline A) (\partial_m A \partial^m A + \partial_m \overline A \partial^m \overline A) - q(A, \overline A) \partial_m A \partial^m \overline A \\
&  - 9 f^2 (g'')^2[ (A + \overline A)g'-A \overline{A}g''-g- 3 ]^{-2} + \text{terms with four or more vectors} \,  ,
\end{aligned}
\ee
where 
\begin{align}
\label{p(A)}
p(A,\overline{A})= &3 (g'')^2 A \overline{A} [(A+\overline{A})g'-g-3]^{-2} - 3 g'' [ (A + \overline A)g'-g- 3 ]^{-1} \nonumber \\
& + 12 g'' [ (A + \overline A)g'-g- 3 ]^{-1}  [ (A + \overline A)g'-A \overline{A}g''-g- 3 ]^2 \times
\\
& \{ 3 [(A + \overline A)g'-A \overline{A}g''-g- 3 ]^2 + g'' A \overline{A} [2g'(A+\overline{A})-2g - 3 g'' A \overline{A}-6]\}^{-1} \nonumber
\end{align}
and 
\begin{align}
    \label{q(A)}
q(A,\overline{A})= &6 (g'')^2 A \overline{A} [(A+\overline{A})g'-g-3]^{-2} - 3 g'' [ (A + \overline A)g'-g- 3 ]^{-1} \nonumber \\
& + 12 g'' [ (A + \overline A)g'-g- 3 ]^{-1}  [ (A + \overline A)g'-A \overline{A}g''-g- 3 ] \times \nonumber \\ &[ (A + \overline A)g'- 2 A \overline{A}g''-g- 3 ]\times
\\
& \{ 3 [(A + \overline A)g'-A \overline{A}g''-g- 3 ]^2 + g'' A \overline{A} [2g'(A+\overline{A})-2g - 3 g'' A \overline{A}-6]\}^{-1}. \nonumber
\end{align}
If we write $A=A_1+iA_2$, where $A_1$ and $A_2$ are real scalar fields, we obtain from \eqref{Llocalonshell} the scalar kinetic terms 
\begin{equation}
    e^{-1} \mathcal{L}_{\text{kin, scalars}} = - (q(A,\overline{A})+2p(A,\overline{A})) \partial_m A_1 \partial^m A_1 - (q(A,\overline{A})-2p(A,\overline{A})) \partial_m A_2 \partial^m A_2
\end{equation}
It is clearly not possible to deduce the signs of the kinetic terms for the real scalars $A_1$ and $A_2$ for arbitrary function $g$, so let us consider the case where
\begin{equation}
g(\Sigma+\bar\Sigma)=  (\Sigma + \bar\Sigma)^2,
\end{equation}
which has been discussed without the higher derivative term \eqref{ep1} in  section 2. 
We then have
\begin{equation}
\label{g,g'.g''}
g=(A+\overline{A})^2 \, , \  g'=2(A+\overline{A}) \, , \ g''=2   \, , 
\end{equation}
from which we find 
\begin{equation}
q(A,\overline{A})+2p (A,\overline{A}) < 0    
\end{equation}
for each value of $A$ satisfying 
\begin{equation}
(A + \overline A)^2 < 3    \, . 
\end{equation}
The last constraint arises from the requirement that the argument of the square root in \eqref{WeylReA1} 
with $g$ and $g'$ given by \eqref{g,g'.g''} be positive. 
Therefore, $A_1$ is always a ghost in this special case.

\section{Conclusions}

In this work we have shown by explicit construction that there exist new methods for local supersymmetry breaking 
with properties clearly distinct from the typical behaviour of D-term or F-term breaking. 
The most striking property of our findings is that the vacuum with broken supersymmetry never contains a 
complete goldstino multiplet and is always in the non-linear phase, 
a property that brings to mind the brane-supersymmetry-breaking scenaria \cite{Antoniadis:1999xk,Angelantonj:1999ms,Dudas:2000nv,Pradisi:2001yv}, 
and their 4D N=1 incarnations \cite{Kachru:2003aw,Cribiori:2019hod}. 
We have found that the new vacua, which can be perturbatively stable de Sitter, 
always come together with a supersymmetric vacuum but for the moment we do not have a tunneling procedure that connects the two vacua, 
as for example can happen for the 3-form goldstino vacua \cite{Farakos:2020wfc}. 
We believe that one of the most important questions, apart from the possible string theory origin of such models, 
is if they have a decay channel to supersymmetric theories or if they are isolated points. 
The latter would mean they are not necessarily near the weak coupling regions of string theory and so their existence 
is on one hand not easily verifiable but on the other hand it means that they are not a priori 
in direct conflict with string theory \cite{Danielsson:2018ztv}. 
In addition, 
the fact that we can find stable de Sitter in 4D N=1 should come as no surprise as it is known that not all 
EFTs have a direct relation to string theory, 
but one should also keep in mind that for the moment we only have strong constraints on 
the weakly coupled/well controlled parameter regions \cite{Andriot:2019wrs,Andriot:2020vlg}.

\section*{Acknowledgements} 

This work is supported by the PEVE-2020 NTUA programme for basic research ``Off-shell supergravity'' with project  number $65228100$. 
FF is supported by the STARS grant SUGRA-MAX.

\appendix

\section{Fermionic formulas}

In this appendix we present some lengthy formulas for the benefit of the reader that we made use of in the main text. 

The supergravity transformation of the auxiliary fermionic component field $\chi_\alpha$ of a complex linear superfield $\Sigma$ reads 
\begin{align}
\nonumber \delta \chi_\alpha = & -2i ( \sigma^b \overline\sigma^a \zeta )_\alpha \hat{D}_a {\overline P}_b - i ( \sigma^a \overline\sigma^b \zeta  )_\alpha \hat{D}_a {\overline P}_b - 4 ( \sigma^b \overline\sigma^a \zeta )_\alpha \hat{D}_a \hat{D}_b \overline A \\ \nonumber & - \frac{8}{3} \zeta_\alpha \overline M \overline F - ( \sigma^m \overline \sigma^n \zeta )_\alpha b_m {\overline P}_n - \frac{1}{3} ( \sigma^n \overline \sigma^m \zeta )_\alpha b_m {\overline P}_n + \frac{2}{3} \zeta_\alpha \overline A R \\
 \nonumber & - \frac{8i}{3} \zeta_\alpha b^m \hat{D}_m \overline A - \frac{4i}{3} \zeta_\alpha \overline A {e_a}^m \mathcal{D}_m b^a - \frac{4}{9} \zeta_\alpha \overline A b_m b^m  \\
 \nonumber & + \frac{2}{3} \overline A b^n \zeta_\alpha \left( \psi_n \sigma^m \overline\psi_m \right) + \frac{2}{3} \overline A b^n \left(\zeta \sigma^m \overline\psi_n \right) \psi_{m \alpha} + \frac{1}{3} \overline A b^n \left( \zeta \psi_m \right) \left( \sigma^m \overline\psi_n \right)_\alpha \\
 \nonumber & - \frac{2i \sqrt{2}}{3} \overline M \left( \zeta \psi_m \right) \left( \sigma^m \overline\lambda \right)_\alpha - \frac{5i \sqrt{2}}{4} b^m \zeta_\alpha \left( \overline\lambda \overline\psi_m \right) - \frac{i}{\sqrt{2}} b_m \zeta_\alpha \left( \overline\lambda \overline\sigma^m \sigma^n \overline\psi_n\right) \\
 \nonumber & + \frac{i\sqrt{2}}{6} b_m \left( \sigma^{mn} \zeta \right)_\alpha \left( \overline\lambda \overline\psi_n \right) - \frac{i \sqrt{2}}{12} b_m \left( \sigma^n \overline \lambda \right)_\alpha \left(\zeta \sigma^m \overline\psi_n \right) - \frac{i \sqrt{2}}{3} b^m \zeta_\alpha \left( \rho \psi_m \right)\\
 \nonumber & + \frac{i \sqrt{2}}{3} b^m \psi_{m \alpha} \left( \zeta \rho\right) + \frac{i}{\sqrt{2}} b^m \rho_\alpha \left(  \zeta \psi_m \right) + \frac{i \sqrt{2}}{3} b_m \left(\sigma^{mn} \rho \right)_\alpha \left( \zeta \psi_n \right)\\
  & + 2 \sqrt{2} \zeta_\alpha \left( \overline \lambda \overline\sigma^{mn} \overline\psi_{mn} \right) + 2 \sqrt{2} \left( \sigma^{mn} \zeta \right)_\alpha \left( \overline\lambda \overline\psi_{mn} \right) + \frac{4 \sqrt{2}}{3} \rho_\alpha \left( \zeta \sigma^{mn} \psi_{mn} \right)\\
 \nonumber & + \frac{2 \sqrt{2}}{3} \left( \sigma^{mn} \zeta\right)_\alpha \left( \rho \psi_{mn} \right) + \frac{2 \sqrt{2}}{3} \psi_{m n \alpha} \left( \zeta \sigma^{mn} \rho \right) - \frac{2 \sqrt{2}}{3} \left(  \sigma^{mn} \rho \right)_\alpha \left( \zeta \psi_{mn} \right) \\
 \nonumber & - \frac{2i}{3} \overline A \zeta_\alpha \left( \psi^m \sigma^n \overline\psi_{mn} \right) + \frac{2i}{3} \overline A \psi_{k \alpha} \left( \zeta \sigma^k \overline\sigma^{mn}  \overline\psi_{mn} \right)
 + \frac{1}{2} \overline A \zeta_\alpha \epsilon^{klmn} \left( \psi_k \sigma_l \overline\psi_{mn} + \frac{1}{3}  \overline\psi_k \overline\sigma_l \psi_{mn}\right) \\
 \nonumber & - 2 i ( \sigma^m \overline \zeta )_\alpha \hat{D}_m \overline F + \frac{1}{3}  ( \sigma^m \overline \zeta )_\alpha M \overline P_m - \frac{4}{3}  ( \sigma^m \overline \zeta )_\alpha b_m \overline F \\
 \nonumber & + \frac{i \sqrt{2}}{3} M \left( \overline\zeta \overline\psi_m \right) \left( \sigma^m  \overline\lambda \right)_\alpha +  \frac{i \sqrt{2}}{3} M \left( \overline\zeta \overline\lambda \right) \left( \sigma^m \overline\psi_m \right)_\alpha + \frac{i \sqrt{2}}{3} M \left( \sigma^m \overline\zeta \right)_\alpha \left( \overline\psi_m \overline\lambda  \right) \\
 \nonumber & + \frac{i \sqrt{2}}{6} b_m \left( \sigma^m \overline \zeta \right)_\alpha \left( \psi_n \sigma^n \overline\lambda \right) + \frac{i \sqrt{2}}{6} b_m \left( \sigma^m \overline \lambda \right)_\alpha \left( \overline\zeta \overline\sigma^n \psi_n \right) - \frac{i \sqrt{2}}{3} b^m \psi_{m \alpha} \left( \overline \zeta \overline \lambda \right) \\
 \nonumber & - \frac{i \sqrt{2}}{6} b_m \psi_{n \alpha} \left( \overline\lambda \overline\sigma^m \sigma^n \overline\zeta \right) - \frac{i \sqrt{2}}{4} b_m \left( \sigma^n \overline \zeta \right)_\alpha \left( \psi_n \sigma^m \overline\lambda \right) - \sqrt{2} \, \psi_{m n \alpha} (\overline\zeta \overline\sigma^{mn} \overline\lambda) \\
 \nonumber & - \frac{\sqrt{2}}{3} \left( \sigma^{mn} \psi_{mn} \right)_\alpha \left( \overline\zeta \overline\lambda \right) .  \end{align}

Also, up to terms that are at least quartic in the fermions, the component form of \eqref{ep1} is given by
\label{ep2}
\begin{align}
\nonumber e^{-1}{\cal L}_\text{HD}=&\frac{1}{64  f^2}\left( P_m P^m \overline{P}_n \overline{P}^n 
-8P_m\overline{P}^mF\overline{F}+16   F^2\overline{F}^2\right) \\ \nonumber
&+ \frac{1}{48 f^2} \left( \overline{M} P_m P^m \overline{\lambda} \overline{\lambda} + M \overline{P}_m \overline{P}^m \lambda \lambda \right) + \frac{1}{12 f^2}  \left( M F \overline{P}_m + \overline{M} \overline{F} P_m \right)\lambda \sigma^m \overline{\lambda} \\ \nonumber
& + \frac{1}{12f^2} \left( \overline{M} \overline{F}^2 \lambda \lambda + M F^2 \overline{\lambda} \overline{\lambda}  \right) - \frac{3i \sqrt{2}}{128 f^2} \left( \overline{P}_m \overline{P}^m P_n \psi^n \lambda - P_m P^m \overline{P}_n \overline{\psi}^n \overline{\lambda}\right)\\ \nonumber
& - \frac{i \sqrt{2}}{64 f^2} P_m \overline{P}^m \left( P_n \overline{\psi}^n \overline{\lambda} - \overline{P}_n \psi^n \lambda \right) + \frac{i \sqrt{2}}{64 f^2} \left( \overline{P}_k \overline{P}^k P_m \psi_n \sigma^{mn} \lambda -  P_k
P^k \overline{P}_m \overline{\psi}_n \overline{\sigma}^{mn} \overline{\lambda}\right) \\ \nonumber
& - \frac{i \sqrt{2}}{32 f^2}  \left( \overline{P}_k \overline{P}_m P_n \psi^k \sigma^{mn} \lambda - P_k P_m \overline{P}_n \overline{\psi}^k \overline{\sigma}^{mn} \overline{\lambda}\right) \\ \nonumber
&- \frac{i \sqrt{2}}{32 f^2} F \overline{F} \left( \overline{P}_m \psi_n \sigma^n \overline{\sigma}^m \lambda - P_m \overline{\psi}_n \overline{\sigma}^n \sigma^m \overline{\lambda} \right) + \frac{i \sqrt{2}}{32 f^2} P_m \overline{P}^m \left( F \psi_n \sigma^n \overline{\lambda} + \overline{F} \overline{\psi}_n \overline{\sigma}^n \lambda \right) \\ \nonumber
&- \frac{i \sqrt{2}}{32 f^2} \left( F P_m \overline{P}_n \psi^n \sigma^m \overline{\lambda} + \overline{F} \overline{P}_m P_n \overline{\psi}^n \overline{\sigma}^m \lambda \right) - \frac{1}{24 f^2} R \left( A F \overline{\lambda} \overline{\lambda} + \overline{A} \overline{F} \lambda \lambda \right) \\ \nonumber
&+ \frac{1}{36 f^2} b_m b^m \left( A F \overline{\lambda} \overline{\lambda} + \overline{A} \overline{F} \lambda \lambda \right) + \frac{1}{32 f^2} b_m \left( \overline{F} \overline{P}^m \lambda \lambda + F P^m \overline{\lambda} \overline{\lambda}\right) \\ \nonumber
& - \frac{1}{48 f^2} R \,  \left( A \overline{P}_m + \overline{A} P_m \right)\lambda \sigma^m \overline {\lambda} + \frac{1}{72 f^2} b_m b^m  \left( A \overline{P}_n + \overline{A} P_n \right)\lambda \sigma^n \overline{\lambda} \\ \nonumber
& - \frac{1}{96 f^2} b^m  \left( P_m \overline{P}_n + \overline{P}_m P_n\right)\lambda \sigma^n \overline{\lambda} - \frac{1}{48 f^2} P_m \overline{P}^m b_n \lambda \sigma^n \overline{\lambda} + \frac{i}{48 f^2} \epsilon^{klmn} P_k \overline{P}_l b_m  \lambda \sigma_n \overline{\lambda} \\ \nonumber
& + \frac{i \sqrt{2}}{48 f^2} b_m \left( A F \overline{P}_n \psi^m \sigma^n \overline{\lambda} + \overline{A} \overline{F} P_n \overline{\psi}^m \overline{\sigma}^n \lambda \right) + \frac{i \sqrt{2}}{96 f^2} b^m \left( A \overline{P}_n \overline{P}^n \psi_m \lambda -  \overline{A} P_n P^n \overline{\psi}_m \overline{\lambda}  \right) \\ \nonumber
& + \frac{1}{48 f^2} b_m \left( F \overline{P}_n \overline{\rho} \overline{\sigma}^m \sigma^n \overline{\lambda} + \overline{F} P_n \rho \sigma^m \overline{\sigma}^n \lambda \right) - \frac{1}{96 f^2} b_m \left( P_n P^n \rho \sigma^m \overline{\lambda} - \overline{P}_n \overline{P}^n \overline{\rho} \overline{\sigma}^m \lambda \right) \\ \nonumber
& - \frac{\sqrt{2}}{32 f^2} \left( F P_m \chi \sigma^m \overline{\lambda} - \overline{F} \overline{P}_m \overline{\chi} \overline{\sigma}^m \lambda \right) - \frac{\sqrt{2}}{64 f^2} \left( P_m \overline{P}_n \overline{\chi} \overline{\sigma}^m \sigma^n \overline{\lambda} + \overline{P}_m P_n \chi \sigma^m \overline{\sigma}^n \lambda \right) \\ 
&  - \frac{\sqrt{2}}{8 f^2} F \overline{F} \left( \chi \lambda + \overline{\chi} \overline{\lambda} \right) - \frac{1}{4 f^2} {e_a}^m \left[ \mathcal{D}_m\left(e^{an} \partial_n A\right) F \overline{\lambda} \overline{\lambda} +  \mathcal{D}_m\left(e^{an} \partial_n \overline{A}\right) \overline{F} \lambda \lambda \right] \\ \nonumber
& - \frac{i}{12 f^2} {e_a}^m \left(\mathcal{D}_m b^a\right) \left( A F \overline{\lambda} \overline{\lambda} - \overline{A} \overline{F} \lambda \lambda \right)
 - \frac{i}{6 f^2} b^m \left[ \left(\partial_m A\right) F \overline{\lambda} \overline{\lambda} - \left(\partial_m \overline{A}\right) \overline{F} \lambda \lambda \right]\\ \nonumber
& - \frac{i}{8f^2} \left( \overline{F} \overline{P}_m \lambda \sigma^{mn}  \mathcal{D}_n \lambda - F P_m \overline{\lambda} \overline{\sigma}^{mn} \mathcal{D}_n \overline{\lambda} \right) - \frac{3i}{16 f^2} {e_a}^m \left[
\overline{F} \left(\mathcal{D}_m \overline{P}^a\right) \lambda \lambda - F \left(\mathcal{D}_m P^a\right) \overline{\lambda} \overline{\lambda} \right] \\ \nonumber
& + \frac{i}{16 f^2} \left[ \overline{P}^m \left(\partial_m \overline{F}\right) \lambda \lambda - P^m \left(\partial_m F\right) \overline{\lambda} \overline{\lambda} \right] \nonumber
 - \frac{i}{16 f^2} \left( \overline{F} \overline{P}^m \lambda \mathcal{D}_m \lambda - F P^m \overline{\lambda} \mathcal{D}_m \overline{\lambda} \right) \\ \nonumber
& - \frac{i}{24 f^2} {e_a}^m \left(\mathcal{D}_m b^a\right)  \left( A \overline{P}_n - \overline{A} P_n \right)\lambda \sigma^n \overline{\lambda}  - \frac{i}{12 f^2} b^m  \left( \overline{P}_n \partial_m A -P_n \partial_m \overline{A} \right)  \lambda \sigma^n \overline{\lambda}\\ \nonumber
& - \frac{1}{8 f^2} {e_a}^m  \left[ P_k
\mathcal{D}_m(e^{an} \partial_n \overline{A}) + \overline{P}_k \mathcal{D}_m\left(e^{an} \partial_n A\right) \right] \lambda \sigma^k \overline{\lambda}\\ \nonumber
& - \frac{i}{8 f^2} \left(\overline{F} P_m \lambda \sigma^m \overline{\sigma}^n \mathcal{D}_n \rho - F \overline{P}_m \overline{\lambda} \overline{\sigma}^m \sigma^n  \mathcal{D}_n \overline{\rho} \right) \\ \nonumber
& - \frac{i}{16 f^2} \left( P_m P^m \overline{\lambda} \overline{\sigma}^n \mathcal{D}_n \rho + \overline{P}_m \overline{P}^m \lambda \sigma^n \mathcal{D}_n \overline{\rho} \right) - \frac{i}{8 f^2}  \left( \overline{F} \partial_m F - F \partial_m \overline{F} \right) \lambda \sigma^m \overline{\lambda}\\ \nonumber
&- \frac{i}{8 f^2} F \overline{F} \left( \lambda \sigma^m \mathcal{D}_m \overline{\lambda} + \overline{\lambda} \overline{\sigma}^m \mathcal{D}_m \lambda \right) - \frac{i}{32 f^2}   \left( P_a \mathcal{D}_m \overline{P}^a - \overline{P}_a \mathcal{D}_m P^a \right)\lambda \sigma^m \overline{\lambda} \\ \nonumber
& - \frac{i}{32 f^2}  \left( \overline{P}^m \mathcal{D}_m P^a - P^m \mathcal{D}_m \overline{P}^a \right)\lambda \sigma_a \overline{\lambda} - \frac{3i}{32 f^2} {e_a}^m  \left( P^n \mathcal{D}_m \overline{P}^a - \overline{P}^n \mathcal{D}_m P^a  \right) \lambda \sigma_n \overline{\lambda}\\ \nonumber
& + \frac{i}{16 f^2} P^m \overline{P}^n \left( \lambda \sigma_m \mathcal{D}_n \overline{\lambda} + \overline{\lambda} \overline{\sigma}_n \mathcal{D}_m \lambda \right) - \frac{1}{32 f^2} \epsilon^{abcd} {e_a}^m  \left( \overline{P}_c \mathcal{D}_m P_d + P_c \mathcal{D}_m \overline{P}_d \right) \lambda \sigma_b \overline{\lambda}\\ \nonumber
& - \frac{\sqrt{2}}{48 f^2} \left( A F \overline{P}_k \overline{\lambda} \overline{\sigma}^k \sigma^m \overline{\sigma}^n \psi_{mn} - \overline{A} \overline{F} P_k \lambda \sigma^k \overline{\sigma}^m \sigma^n \overline{\psi}_{mn} \right) \\ \nonumber
& - \frac{\sqrt{2}}{16 f^2} \left( F \overline{P}_m \psi_n \sigma^k \overline{\sigma}^n \sigma^m \overline{\lambda} \, \partial_k A - \overline{F} P_m \overline{\psi}_n \overline{\sigma}^k \sigma^n \overline{\sigma}^m \lambda \, \partial_{k} \overline{A} \right) \\ \nonumber
& - \frac{\sqrt{2}}{32 f^2} \left( P_m P^m \overline{\psi}_n \overline{\sigma}^k \sigma^n \overline{\lambda} \partial_k \overline{A} + \overline{P}_m \overline{P}^m \psi_n \sigma^k \overline{\sigma}^n \lambda \partial_k A \right) \\ \nonumber
& + \frac{\sqrt{2}}{48 f^2} \left( A \overline{P}_k \overline{P}^k \lambda \sigma^{mn} \psi_{mn} + \overline{A} P_k P^k \overline\lambda \overline{\sigma}^{mn} \overline{\psi}_{mn} \right) + \frac{7}{24 f^2} F \overline{F} b_m \lambda \sigma^m \overline\lambda \, . 
\end{align}

\end{document}